\documentclass[linenumbers, preprint2, tighten]{jaa}

\usepackage{graphicx}
\usepackage{xcolor, soul}

\begin{document}\sloppy

\title{Spectral properties of the Be/X-ray pulsar 2S 1553-542 during Type II outbursts}


\author{Binay Rai\textsuperscript{1}, Biswajit Paul\textsuperscript{2}, Mohammed Tobrej\textsuperscript{1}, Manoj Ghising\textsuperscript{1}, Ruchi Tamang\textsuperscript{1} and Bikash Chandra Paul\textsuperscript{1}}
\affilOne{\textsuperscript{1}Department of Physics, North Bengal University,Siliguri, Darjeeling, WB, 734013, India\\}
\affilTwo{\textsuperscript{2}ICARD, Depatment of Physics, North Bengal University, Siliguri, Darjeeling, WB, 734013, India}


\twocolumn[{

\maketitle

\corres{bcpaul@associates.iucaa.in}


\begin{abstract}
We present an extended study of the Be/X-ray pulsar 2S 1553-542 during its Type II outbursts. We have incorporated \emph{NICER, Swift-XRT, RXTE-PCA, \emph{NuSTAR} and FERMI} observations to carry out the detailed phase and time resolved spectral analysis of the source. We have summarized the evidence of variability of the cyclotron feature observed in the X-ray continuum of the source with respect to the pulse phases of the pulsar by using the recent \emph{NuSTAR} observation of 2021 outburst of the source. The time resolved spectral analysis has been performed by considering \emph{RXTE} observations of the 2008 outburst of the pulsar. The hardness intensity diagram (HID) has been obtained using 2008 observations in which the intensity follows distinct branches with respect to the hardness ratio. Diagonal branch is observed in the high intensity state whereas the horizontal branch corresponds to the low intensity state. The transition from the diagonal to horizontal branch occurs at the luminosity of $(4.88\pm0.24)\;\times\;10^{37}\;erg\;s^{-1}$. The photon-index exhibits a weak positive correlation with flux along the diagonal branch and negative correlation along the horizontal branch. The existence of two different diagonal and horizontal branches further reflects the possibility of two different accretion states separated by the critical luminosity. The spin-up rate during the outburst phase is found to depend on the flux and is found to increase with an increase in flux.
\end{abstract}


}]


\doinum{12.3456/s78910-011-012-3}
\artcitid{\#\#\#\#}
\volnum{000}
\year{0000}
\pgrange{1--}
\setcounter{page}{1}
\lp{1}

 \section{Introduction}

\textcolor{black}{2S 1553-542 is a Be/X-ray pulsar that is transient and detectable during outburst. In this type of X-ray pulsar, the companion of the neutron star is a Be star.} \textcolor{black}{These pulsars are visible only during bright outbursts.} A typical Be star is a non-supergiant B-type star having luminosity class between III-V that exhibits spectral lines \textcolor{black}{(e.g. Porter 2003 ; Rivinius et al. 2013)}. \textcolor{black}{A Be star consists of a B type star with an equatorial disc}. During a periastron, the neutron star may pass close to \textcolor{black}{the} disc or even pass through it, as a result of which a large amount of matter gets accreted onto the neutron star resulting in Type I outburst. \textcolor{black}{This kind of outbursts are frequent and the luminosity may reach up to $10^{37} erg \;s^{-1}$} (for review on Be/X-ray binaries see Reig 2011). \textcolor{black}{Along with Type I outbursts Be/X-ray sources also exhibit Type II outbursts which are more luminous and rare than Type I outbursts (Reig 2011).}

 In black-hole X-ray binaries (BXBs) and low-mass X-ray binaries (LMXBs) the definition of the source states is very useful in understanding the different phenomena shown by these sources in X-ray band (Hasinger \& van der Klis 1989; Homan \& Belloni 2005; Belloni 2010). These states are defined in terms of spectral appearance or X-ray variability which are associated with the specific well-defined position in the hardness intensity diagram (HID) or colour-colour diagram. Reig \& Nespoli (2013) implemented this method to study the spectral and X-ray variability of Be/X-ray binaries during a giant outburst. \textcolor{black}{It was observed} that the hardness-intensity diagram of Be/X-ray pulsars traces two different branches - the horizontal branch in the low-intensity state and the diagonal branch in  the high-intensity state when the luminosity of the source exceeds a certain limit.  \textcolor{black}{The critical luminosity $L_{c}$ which marks the transition in the accretion schemes reflects the transition between these two branches (Basko \& Sunyaev 1976). The strong magnetic field of the neutron star in accretion powered pulsars hinders the accretion flow at some distance from the surface of the neutron star which forces the accreted matter to flow down to the polar caps. This creates hot spots and an accretion column. The two accretion regimes i.e., sub-critical and super-critical regime  are characterized by different ways in which the radiation pressure of the emitting plasma is capable of decelerating the accretion flow.  At higher luminosities (the super-critical regime), the radiation pressure is high and the braking of the accreting matter flow occurs due to the interaction with photons. A radiation-dominated shock is responsible for stopping the flow at a certain distance from the neutron star surface (Davidson \& Ostriker 1973; Basko \& Sunyaev 1976; Lyubarskii \& Syunyaev 1982). However, in the sub-critical regime, the pressure of the radiation-dominated shock is not enough to cease the flow. It continues to flow to the surface of the neutron star where it undergoes a deceleration by multiple Coulomb scattering with thermal electrons and nuclear collisions with atmospheric protons (Burnard et al. 1991; Harding 1994). For very low luminosities, the radiation dominated shock does not exist. Reig \& Nespoli (2013) proposed that the transition from the sub-critical to the super-
critical regime, reflects two different branches in their HID. The horizontal branch (HB) pattern is observed in the sub-critical regime while the diagonal branch (DB) pattern is observed in the super-critical regime.}

 2S 1553-542 was discovered in 1975 by \textit{SAS-3} observatory (Walter 1976). The pulsating nature of the source was discovered later by Kelley (1982). The pulse period of the source is about 9.3 s and the binary orbital period is 30.6 days (\textcolor{black}{Kelley 1982)}. \textcolor{black}{The second time the pulsar was detected} in 2007 (Krimm 2007) during a type II outburst. The hard X-ray flares from the source \textcolor{black}{were} observed by Pahari \& Pal (2012). They also observed that with the decay of the flare the pulse fraction also decreases but its value changes discontinuously with energy. \textcolor{black}{The third active period in X-rays} of 2S 1553-542 was observed in 2015 type II outburst. Tsygankov et al. (2016) discovered the cyclotron absorption feature in the source spectrum using the observation from the \textit{NuSTAR} observatory. The accuracy of the binary parameters was also improved by these authors. The estimated distance to the source is $\sim$ 20 kpc (Tsygankov et al. 2016). Lutovinov et al. (2016) reported the study of source in near-infrared to X-ray wavelength. Their study \textcolor{black}{suggests} that the system shows moderate NIR excess which is expected from a typical Be star, as it is due to presence of circumstellar disc around the star. The comparison of spectral energy distribution with the known Be/X-ray binaries led Lutovinov et al. (2016) to estimate the spectral type of companion to be B1-2V corresponding to which the distance to the source is $>$ 15 kpc. The \textcolor{black}{most} recent outburst of the pulsar was discovered in the first week of January 2021 by Fermi/GBM (Jenke et al. 2021). The follow-up observations were made using the observatories like \textit{Swift, MAXI, INTEGRAL, NuSTAR} etc. The observation of the pulsar by \textit{NuSTAR} (Malacaria 2021) revealed that the presence of cyclotron lines \textcolor{black}{at 27 keV} and the flux in 3-79 keV energy range to be $\sim$ 1.6$\times$10$^{-9}$ erg cm$^{-2}$ s$^{-1}$. Malacaria et al. (2022) revealed that the cyclotron line energy of the pulsar is luminosity dependent. The luminosity and the duration of the outburst suggest that the recent 2021 outburst is of type II.

\textcolor{black}{The study of spectral features of X-ray pulsars provides information on different absorption and emission properties. In particular, one of the widely studied characteristic absorption in the continuum spectra of X-ray pulsars is Cyclotron Resonance Scattering Feature (CRSF). The origin of the CRSF feature takes place at the magnetic poles of a neutron star where the strength of the magnetic field is strong. During accretion, the matter accreted by the neutron star falls onto the magnetic poles. As a result, the kinetic energy of the infalling matter is converted into heat and radiation. The presence of a strong magnetic field at poles leads the electrons present (perpendicular to the magnetic field) on the infalling matter to get quantized into discrete energy levels. These levels are also known as Landau levels (Sch{\"o}nherr et al. 2007). Photons having energies close to these Landau levels are scattered by those quantized electrons leading to the formation of a cyclotron line in the spectra. Knowing cyclotron line energy of CRSF feature, one can estimate the magnetic field strength on the surface of a neutron star by;}

 \begin{equation} B=\frac{E_{cyc}(1+z)}{11.57} 10^{12}\;G \end{equation} 

\textcolor{black}{where $E_{cyc}$ is the cyclotron line energy of the CRSF in units of keV, and z is the Gravitational red-shift.}

  As the average spectroscopic results have been reported by Malacaria et al. (2022), it would be interesting to verify the dependence of the spectral parameters on changing viewing angle of the neutron star. The authors have also shown the variation of the cyclotron line using 2015 and 2021 observations of the pulsar. In this paper, we have performed phase-resolved spectral of the 2S 1553-542 using the \emph{NuSTAR} observation of it during 2021 outbursts. Using the \textit{RXTE} observations from the earlier 2008 type II outburst we have tried to show a correlation between the hardness intensity diagram with the spectral properties of the source. At last, we have shown the spin evolution of the pulsar with time and tried to study the dependence of the spin-up rate on the X-ray luminosity using the spin history of the pulsar provided by \emph{FERMI}-GAPP team. We have used the data from the multiple X-ray observatories in our study, this has been discussed in the proceeding section.   

\section{Observation and Data reduction}
 We have used data from Swift/XRT, \emph{NICER}, \emph{RXTE}, \emph{NuSTAR}, and FERMI GBM observations to present detailed coverage of the source. The data has been processed using heasoft v6.29 using the suitable calibration files related to the different instruments of different missions.
\subsection{Swift}
 The Neil Gehrels Swift observatory (Gehrels et al. 2004) consists of three different instruments - BAT, XRT, and UVIT - out of which we have used observations made using XRT (X-ray telescope) operating in a soft X-ray range (0.5-10 keV). For \textit{Swift}-XRT data the necessary screening and filtering were done using \textsc{xrtpipeline}. We have used 17 pointed observations of \textit{Swift}-XRT. The specifications related to the observation IDs and corresponding exposure time have been presented in Table 1. All the observations were made in PC mode of XRT. The count rates of the source were greater than 0.5 count s$^{-1}$, so to avoid the pileup effect we excluded the central region of radius 5$^{\prime\prime}$ of the source and considered an annular region of inner and outer radii 5$^{\prime\prime}$ and 25$^{\prime\prime}$ respectively as a source extraction region. A circular region of 20$^{\prime\prime}$ away from the source was taken as the background region. From these regions, the source and background spectra were extracted in \textsc{xselect}. The ancillary response file \textsc{(arf)} was extracted using \textsc{xrtmkarf} and \textsc{rmf} was obtained from the latest calibration database files (CALDB v 20210915). The variation of the MAXI count rate throughout this outburst along with that of Swift is given in Figure 1.
 \subsection{\emph{NICER}}
 The Neutron Star Interior Composition Explorer (\textit{NICER}) is an
X-ray telescope onboard International Space Station (ISS) (Arzoumanian et al. 2014; Gendreau et al. 2016). We have used four \textit{NICER} observations in the study. The standard screening and filtering of \textit{NICER} observational data were done with the help of a tool \textsc{nicerl2} which is a part of \textsc{nicerdas} v9, CALDB v xti20210720. The default values of parameters are used while screening and filtering. The background spectra for each observation was extracted using the background estimator tool \textsc{nibackgen3c50} v7. The ancillary response file and the response matrix file were extracted using \textsc{nicerarf} and \textsc{nicerrmf}. The \textit{NICER} spectra between 0.7-10 keV were only considered for fitting, above 10 keV there will be an additional noise and below 0.7 the spectra are dominated by background (Malacaria et al. 2022). The NICER observations are shown by \textcolor{black}{solid arrows lines} in Figure 1.

\subsection{\emph{RXTE}}

To study the temporal and spectral variability of the pulsar during an outburst we have used 63 pointed Rossi X-ray Timing Explorer (RXTE)/ Proportional Counter Array (PCA) observations of the pulsar during the 2008 outburst. The observation ID and the exposures of the \emph{RXTE} observations are given in Table 1. The data were filtered using the standard criteria - time of passage of SAA greater than 30 minutes, Earth angle of elevation greater than 10 degrees, and point offset less than 0.01 degrees. The light curves and spectra were extracted using STANDARD 2 data of \textit{RXTE}/PCA. The only top layer of the PCU2 data was used while extracting the spectra. The background spectra and light curves were extracted using the tool \textsc{runpcabackest} with the help of a bright background model and PCA SAA history file. The response files for the spectral analysis were generated using \textsc{pcarsp}. 
\textcolor{black}{The RXTE spectrum above 23 keV was background dominated, therefore we restrict our analysis to the 3-23 keV range.} A systematic error of 1.5$\%$ was added to all the spectra. The background correction of the light curves was made using \textsc{lcmath}. The barycentric correction of the light curves was made using \textsc{barycorr}. All spectra were fitted in \textsc{xspec} \; v12.12.1 (Arnaud 1996). 

\subsection{\emph{NuSTAR}}
 
\textcolor{black}{The Nuclear Spectroscopic Telescope Array (\emph{NuSTAR}) mission has two identical co-aligned X-ray telescopes focusing X-ray photons onto two Focal Plane Modules FPMA and FPMB consisting
of a pixelated solid-state detector (CdZnTe) (Harrison et al. 2013). It operates in the (3-79) keV energy range. The source region for the analysis was selected from a circular region of 100$^{\prime\prime}$ and a background region of 100$^{\prime\prime}$ located far enough from the source using astronomical imaging and data visualization application DS9\footnote{https://sites.google.com/cfa.harvard.edu/saoimageds9}. Data reduction was performed with the standard \emph{NuSTAR} Data Analysis Software (NUSTARDAS) v2.0.0 provided under heasoft v6.29 using CALDB v 20210210 following the data analysis manual. Background corrected light curves were generated using ftool LCMATH. Barycentric corrections were done with the \textcolor{black}{BARYCORR} utility. The source spectra obtained were binned following the optimal binning algorithm proposed by Kaastra \& Bleeker (2016). The spectra so obtained were fitted in XSPEC v12.12.0.} \textcolor{black}{The NuSTAR observation details are mentioned in Table 1.}

\subsection{Fermi/GBM}
\textcolor{black}{The Fermi Gamma-Ray Space Telescope studies gamma-ray sources and was launched in 2008. It has a Gamma - ray Burst Monitor (GBM) which is efficient in the energy range of 20 MeV to about 300 GeV (Meegan et al. 2009). We have considered the publicly available pulsar data for our analysis as it provides an estimate of the source flux in the hard band. The spin frequency provided by FERMI GBM Accreting Pulsars Program (GAPP) has been used to obtain the frequency derivative by linear fitting three consecutive frequencies (Serim et al. 2022). The rate of change of frequency is given by the slope of the plot and the error associated with it corresponds to 1-sigma uncertainty in measuring the spin-up rate.}

\section{Analysis and Results}

\begin{table*}
\scalebox{0.6}{
\centering
\begin{tabular}{lccccccccl}
\hline											
Observatory	&	Obs Ids	&	Date of Obs      (in MJD)	&	Exposure (ks)	&	Observatory	&	Obs Ids	&	Date of Obs      (in MJD)	&	Exposure (ks)	\\
\hline															
\emph{RXTE}	&	93426-01-01-00	&	54471.15	&	1.97	&	\emph{RXTE}	&	93426-01-08-05	&	54516.56	&	1.46	\\
\emph{RXTE}	&	93426-01-02-00	&	54475.31	&	2.84	&	\emph{RXTE}	&	93426-01-09-00	&	54518.01	&	2.09	\\
\emph{RXTE}	&	93426-01-02-01	&	54474.83	&	1.45	&	\emph{RXTE}	&	93426-01-09-01	&	54519.69	&	1.46	\\
\emph{RXTE}	&	93426-01-02-02	&	54473.54	&	3.73	&	\emph{RXTE}	&	93426-01-09-02	&	54520.54	&	1.73	\\
\emph{RXTE}	&	93426-01-02-03	&	54474.26	&	1.1	&	\emph{RXTE}	&	93426-01-09-04	&	54522.52	&	3.81	\\
\emph{RXTE}	&	93426-01-03-00	&	54476.54	&	3.53	&	\emph{RXTE}	&	93426-01-10-00	&	54525.39	&	1.08	\\
\emph{RXTE}	&	93426-01-03-01	&	54477.55	&	2.66	&	\emph{RXTE}	&	93426-01-10-02	&	54529.11	&	3.24	\\
\emph{RXTE}	&	93426-01-03-02	&	54478.35	&	1.77	&	\emph{RXTE}	&	93426-01-10-03	&	54530.58	&	1.73	\\
\emph{RXTE}	&	93426-01-03-03	&	54479.54	&	3.14	&	\emph{RXTE}	&	93426-01-11-00	&	54532.03	&	4.17	\\
\emph{RXTE}	&	93426-01-03-04	&	54480.48	&	2.17	&	\emph{RXTE}	&	93426-01-11-01	&	54533.05	&	2.12	\\
\emph{RXTE}	&	93426-01-03-05	&	54480.28	&	1.87	&	\emph{RXTE}	&	93426-01-11-02	&	54534.01	&	1.48	\\
\emph{RXTE}	&	93426-01-03-06	&	54482.2	&	1.64	&	\emph{RXTE}	&	93426-01-11-03	&	54535.02	&	2.27	\\
\emph{RXTE}	&	93426-01-03-07	&	54481.93	&	1.02	&	\emph{RXTE}	&	93426-01-11-04	&	54536	&	4.54	\\
\emph{RXTE}	&	93426-01-04-00	&	54483.58	&	2.13	&	\emph{RXTE}	&	93426-01-11-05	&	54537.12	&	4.05	\\
\emph{RXTE}	&	93426-01-04-01	&	54484.63	&	2.02	&	\emph{RXTE}	&	93426-01-11-07	&	54538.31	&	3.33	\\
\emph{RXTE}	&	93426-01-04-02	&	54485.15	&	1.51	&	\emph{RXTE}	&	93426-01-12-01	&	54540.78	&	5.7	\\
\emph{RXTE}	&	93426-01-04-03	&	54486.87	&	1.96	&	\emph{RXTE}	&	93426-01-12-02	&	54541.71	&	3.4	\\
\emph{RXTE}	&	93426-01-04-04	&	54487.45	&	1.36	&	\emph{RXTE}	&	93426-01-12-04	&	54543.73	&	3.64	\\
\emph{RXTE}	&	93426-01-04-05	&	54488.06	&	2.01	&	\emph{RXTE}	&	93426-01-12-05	&	54544.1	&	2.88	\\
\emph{RXTE}	&	93426-01-04-06	&	54489.03	&	2.3	&	\emph{RXTE}	&	93426-01-12-06	&	54545.69	&	4.01	\\
\emph{RXTE}	&	93426-01-05-00	&	54490.19	&	3.76	&	\emph{NICER}	&	3202030101	&	59268.0781	&	1.43	\\
\emph{RXTE}	&	93426-01-05-02	&	54492.82	&	1.6	&	\emph{NICER}	&	3202030102	&	59271.625	&	2.08	\\
\emph{RXTE}	&	93426-01-05-03	&	54493.92	&	3.53	&	\emph{NICER}	&	3202030103	&	59272.0156	&	0.41	\\
\emph{RXTE}	&	93426-01-05-04	&	54494.84	&	2.89	&	\emph{NICER}	&	3202030104	&	59274.0156	&	0.34	\\
\emph{RXTE}	&	93426-01-05-05	&	54495.89	&	2.95	&	\textit{Swift}	&	00031096003	&	59221.2891	&	1.35	\\
\emph{RXTE}	&	93426-01-05-06	&	54496.67	&	1.97	&	\textit{Swift}	&	00031096004	&	59225.0039	&	1.25	\\
\emph{RXTE}	&	93426-01-06-00	&	54497.98	&	3.52	&	\textit{Swift}	&	00031096005	&	59227.1953	&	1.29	\\
\emph{RXTE}	&	93426-01-06-01	&	54498.96	&	3.51	&	\textit{Swift}	&	00031096006	&	59229.8711	&	0.23	\\
\emph{RXTE}	&	93426-01-06-02	&	54500.01	&	1.7	&	\textit{Swift}	&	00031096007	&	59231.1953	&	0.26	\\
\emph{RXTE}	&	93426-01-06-03	&	54500.92	&	3.51	&	\textit{Swift}	&	00031096008	&	59233.832	&	1.31	\\
\emph{RXTE}	&	93426-01-06-04	&	54501.97	&	3.5	&	\textit{Swift}	&	00031096009	&	59235.3047	&	1.44	\\
\emph{RXTE}	&	93426-01-06-05	&	54502.95	&	3.45	&	\textit{Swift}	&	00031096010	&	59239.2852	&	1.6	\\
\emph{RXTE}	&	93426-01-07-00	&	54504	&	2.75	&	\textit{Swift}	&	00031096012	&	59243.9844	&	1.3	\\
\emph{RXTE}	&	93426-01-07-01	&	54505.96	&	3.51	&	\textit{Swift}	&	00031096014	&	59253.082	&	0.88	\\
\emph{RXTE}	&	93426-01-07-02	&	54506.75	&	3.5	&	\textit{Swift}	&	00031096016	&	59263.332	&	0.95	\\
\emph{RXTE}	&	93426-01-07-03	&	54507.79	&	4.99	&	\textit{Swift}	&	00031096017	&	59268.7656	&	0.85	\\
\emph{RXTE}	&	93426-01-07-04	&	54509.64	&	1.94	&	\textit{Swift}	&	00031096018	&	59273.1484	&	0.71	\\
\emph{RXTE}	&	93426-01-07-05	&	54510.69	&	3.39	&	\textit{Swift}	&	00031096019	&	59278.9844	&	0.96	\\
\emph{RXTE}	&	93426-01-07-06	&	54509.71	&	1.48	&	\textit{Swift}	&	00031096020	&	59283.0352	&	1.04	\\
\emph{RXTE}	&	93426-01-08-01	&	54511.08	&	1.76	&	\textit{Swift}	&	00031096021	&	59285.4141	&	0.72	\\
\emph{RXTE}	&	93426-01-08-02	&	54512.97	&	2.07	&	\textit{Swift}	&	00031096022	&	59287.082	&	0.96	\\
\emph{RXTE}	&	93426-01-08-03	&	54513.69	&	2.09	&	\emph{NuSTAR}	&	90701302002	&	59236.9062	&	28.35	\\
\emph{RXTE}    & 93426-01-08-03 &     54515.92  &  1.23 & & & & \\
\hline
\end{tabular}}
\caption{List of observations used in this paper. The \emph{\emph{RXTE}} observations corresponds to 2008 outburst and rest of the observations belongs to the recent 2021 outburst. }
\end{table*}

\subsection{Pulse period estimation}

\textcolor{black}{In order to estimate the pulse period of the source 2S 1553-542, we considered \emph{NuSTAR} light curves with binning of 0.01 s. Light curves were plotted using ftool \textsc{lcurve}. Using Fast Fourier transform (FFT) on the light curve, we estimated the approximate pulse period. In order to estimate the precise pulse period, we used epoch-folding technique (Davies 1990; Larsson 1996). This method is based on $\chi^{2}$ maximization and hence determined the best period 9.2822$\pm$0.0001s. The error associated with the pulse period is computed using the method given by Boldin et al. (2013).}

\begin{figure*}
\begin{minipage}{0.35\textwidth}
\includegraphics[height=1\columnwidth]{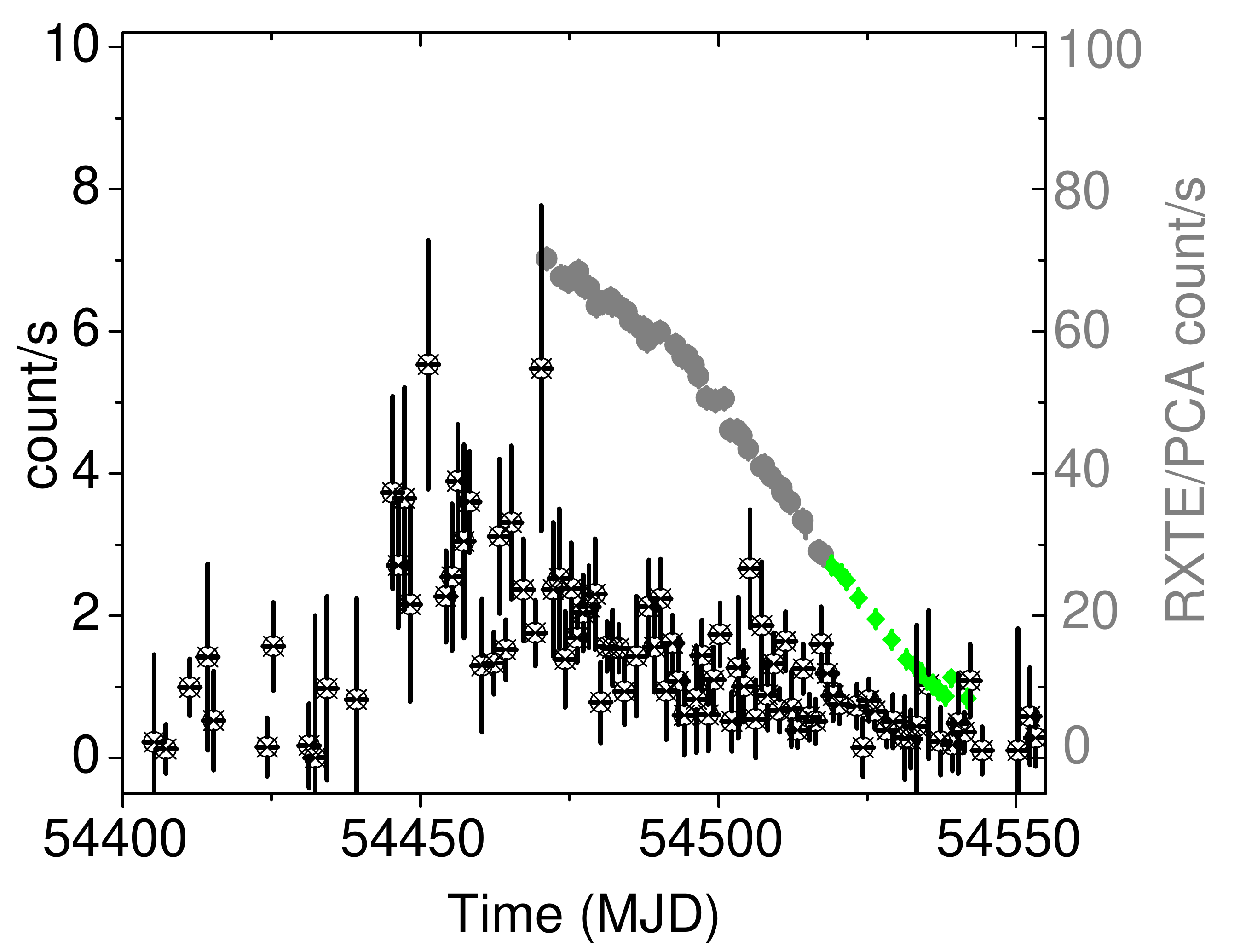}
\end{minipage}
\hspace{0.2\linewidth}
\begin{minipage}{0.35\textwidth}
\includegraphics[height=1.1\columnwidth]{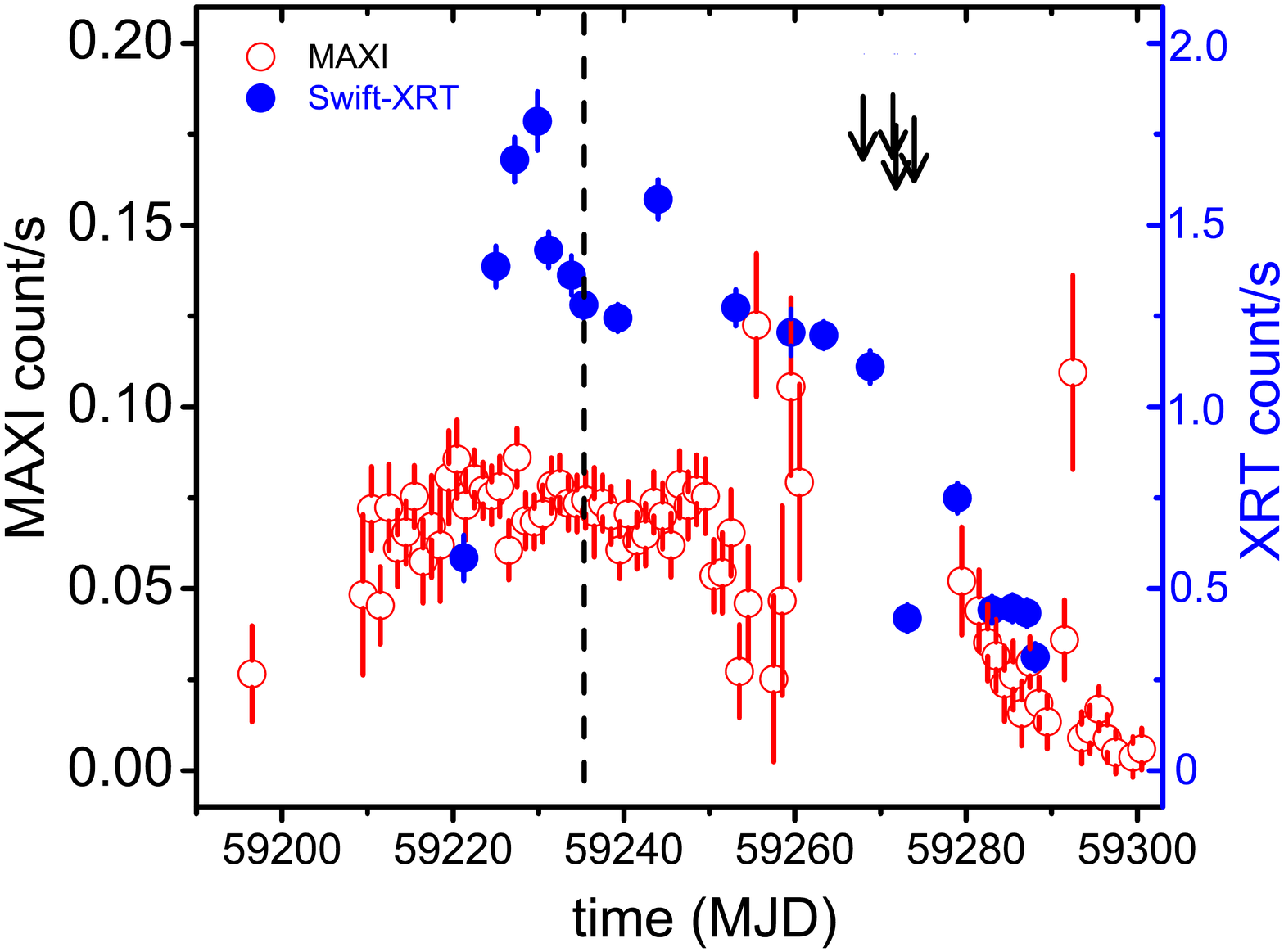}
\end{minipage}
\caption{ \textit{Left panel} - \emph{RXTE}/ASM (black) light curve of the 2008 outburst where grey and green points represents \emph{RXTE}/PCA count rates for DB (Diagonal branch) and HB (Horizontal branch) of the Hardness Intensity Diagram respectively of the same outburst. \textit{Right panel } - Light curve showing the variation of \emph{Swift}-XRT and MAXI count rate of pulsar 2S 1553-542 during the recent outburst in 2021. The solid arrow marked in black represents the time of the \textit{NICER} observations. MAXI and \emph{Swift}-XRT observations have been marked by hollow red circles and filled blue circles respectively. The lone \emph{NuSTAR} observation has been represented by vertical dotted line.}
\end{figure*}

\subsection{Phase resolved spectral analysis}

\begin{table*}
\scalebox{0.8}{
\begin{tabular}{lcccccccccl}
\hline
Parameters		&	&	MODEL I (CUTOFFPL)	&	&	&	MODEL II (HIGHECUT)	&	&	&	MODEL III (COMPTT) 	\\
\hline													
$C_{FPMA}$		&	&	1(fixed)	&	&	&	1(fixed)	&	&	&	1(fixed)	\\
$C_{FPMB}$		&	&	1.008$\pm$0.002	&	&	&	1.007$\pm$0.002	&	&	&	1.002$\pm$0.003	\\
$n_{H}\;(cm^{-2})$		&	&	1.95$\pm$0.22	&	&	&	3.62$\pm$0.30	&	&	&	0.99$\pm$0.31	\\
CompTT ($T_{o}$) (keV)		&	&	-	&	&	&	-	&	&	&	1.01$\pm$0.04	\\
CompTT (kT) (keV)		&	&	-	&	&	&	-	&	&	&	4.59$\pm$0.06	\\
CompTT ($\tau$) (keV)		&	&	-	&	&	&	-	&	&	&	6.72$\pm$0.17	\\
$\Gamma$		&	&	0.23$\pm$0.03	&	&	&	1.12$\pm$0.06	&	&	&		\\
$E_{cut}$ (keV)		&	&	6.84$\pm$0.12	&	&	&	12.03$\pm$0.18	&	&	&		\\
$E_{fold}$ (keV)		&	&	-	&	&	&	8.30$\pm$0.28	&	&	&	-	\\
Fe line (keV)		&	&	6.27$\pm$0.09	&	&	&	6.29$\pm$0.10	&	&	&	6.34$\pm$0.06	\\
$\sigma_{Fe}$ (keV)		&	&	0.45$\pm$0.09	&	&	&	0.48$\pm$0.16	&	&	&	0.50$\pm$0.08	\\
$E_{cyc}$ (keV)		&	&	27.40$\pm$0.44	&	&	&	26.69$\pm$0.76	&	&	&	27.28$\pm$0.47	\\
$\sigma_{cyc}$ (keV)		&	&	5.25$\pm$0.33	&	&	&	7.23$\pm$0.85	&	&	&	7.29$\pm$0.54	\\
$Strength_{cyc}$ (keV)			&	&	6.32$\pm$0.79	&	&	&	7.35$\pm$1.99	&	&	&	12.98$\pm$2.16	\\
Flux ($\times   10^{-9}\;erg\;cm^{-2}\;s^{-1}$)		&	&	1.69$\pm$0.01	&	&	&	1.69$\pm$0.02	&	&	&	1.69$\pm$0.03	\\
$\chi^{2}_{\nu}$ \textcolor{black}{(d.o.f)}		&	&	1.10\textcolor{black}{(1223)}	&	&	&	1.12 \textcolor{black}{(1222)}	&	&	&	1.14\textcolor{black}{(1222)}	\\

\hline
\end{tabular}}
\caption{\textcolor{black}{The above table represents the best-fitted parameters of the source 2S 1553-542 for NuSTAR observation using the continuum model CONSTANT*PHABS*(CUTOFFPL+GA)*GABS, CONSTANT*PHABS*(HIGHECUT*PL+GA)*GABS and CONSTANT*PHABS*(COMPTT+GA)*GABS. $n_{H}$ represents equivalent hydrogen column density for PHABS component in units of $10^(22)$,
$\Gamma$, and $E_{CUT}$ represents Photon Index and CUTOFF energy of CUTOFFPL model. $E_{fold}$ represents folded energy of the HIGHECUT component, Fe and $\sigma_{Fe}$ represents iron line and its equivalent width for the GAUSSIAN component, $T_{o}$, kT and $\tau$ represents the soft comptonization temperature, plasma temperature and strength of the COMPTT component.  $E_{cyc}$, $\sigma_{cyc}$ and $Strength_{cyc}$ represents the Cyclotron line energy, width and the strength of the GABS component. The flux is estimated in the energy range 3-79 keV and errors for each parameters are within 90 \% uncertainty. }}
\end{table*}

In this section, we present an extended overview of spectral analysis of the source by considering the periodic properties of the source using \emph{NuSTAR} observation. To find the best models combination to perform phase-resolved analysis, we fitted the phase-averaged \emph{NuSTAR} FPMA \& FPMB 3-79 keV spectra using different models combinations. The source and background spectra were generated by following the standard procedures as described in Section 2.4. Using appropriate response matrices files (RMF), background, and ancillary response files (ARF), spectral fitting was carried out by XSPEC v12.8.2 package. First, we tried to fit the spectrum of the X-ray pulsar using power-law model modified with a cutoff at higher energies (CUTOFFPL). Therefore, we initially approximated the spectrum with a CUTOFFPL model. The model CONSTANT was used to account for the instrumental uncertainties between FPMA \textcolor{black}{and} FPMB and PHABS to estimate the amount of the photo-electric absorption along the direction of the source. The cross-section used in estimating photo-electric cross-section was \textsf{vern} (Verner et al. 1996) with solar abundance set to \textsf{angr} (Andres \& Grevesse 1989). Some residuals were seen in the form of emission at (6-7) keV energy band and at 25-30 keV in the form of absorption. So, incorporation of a \textsf{GAUSSIAN} (hereafter GA) and \textsf{GABS} model to fit the emission and absorption feature. The result of this fitting (Model I - \textsf{CONSTANT*PHABS*(CUTOFFPL+GA)*GABS)}, yielded a reduced $\chi ^{2}$ value of 1.10. The other model combination used were \textsf{CONSTANT*PHABS*(HIGHECUT*PL+GA)*GABS} (Model II) and \textsf{CONSTANT*PHABS*(COMPTT+GA)*GABS} (Model III). The spectral parameters corresponding to the different continuum models have been presented in Table 2.\textcolor{black}{The fitted spectra using different models combination are shown in Figure 2.} All model combinations fits the \emph{NuSTAR} spectrum very well, \textcolor{black}{however, Model I makes it slightly better.} So we fitted the phase-resolved spectra by Model I. \textcolor{black}{It is worth mentioning that in absence of GABS and GA model the CUTOFFPL model gives reduced $\chi^{2}$ as 2.09 for 1229 dof. With an addition of GA model reduced $\chi^{2}$ decreases to 1.99. The reduced $\chi^{2}$ further decreases to 1.10 for 1223 dof with an inclusion of GABS model, thus justifying the addition GA and GABS model}. \textcolor{black}{The discrepancy in the value  of the photon-index of CUTOFFPL obtained Malacaria et al. (2022) can be due to the fact that they have modeled the soft part of the broadband spectrum (0.5-79) keV of the pulsar by hot blackbody model which is not required here. Also the phase-averaged spectral fitting performed here is in 3-79 keV energy range.} The number of spectral bins is comparatively low in phase-resolved spectra as compared to phase-averaged spectra (reduced signal to noise ratio), due to which we were not able to constraint the iron emission line and hence dropped the \textsf{GA} model.

\begin{figure*}
\begin{minipage}{0.33\textwidth}
\includegraphics[height=1.0\columnwidth, angle=-90]{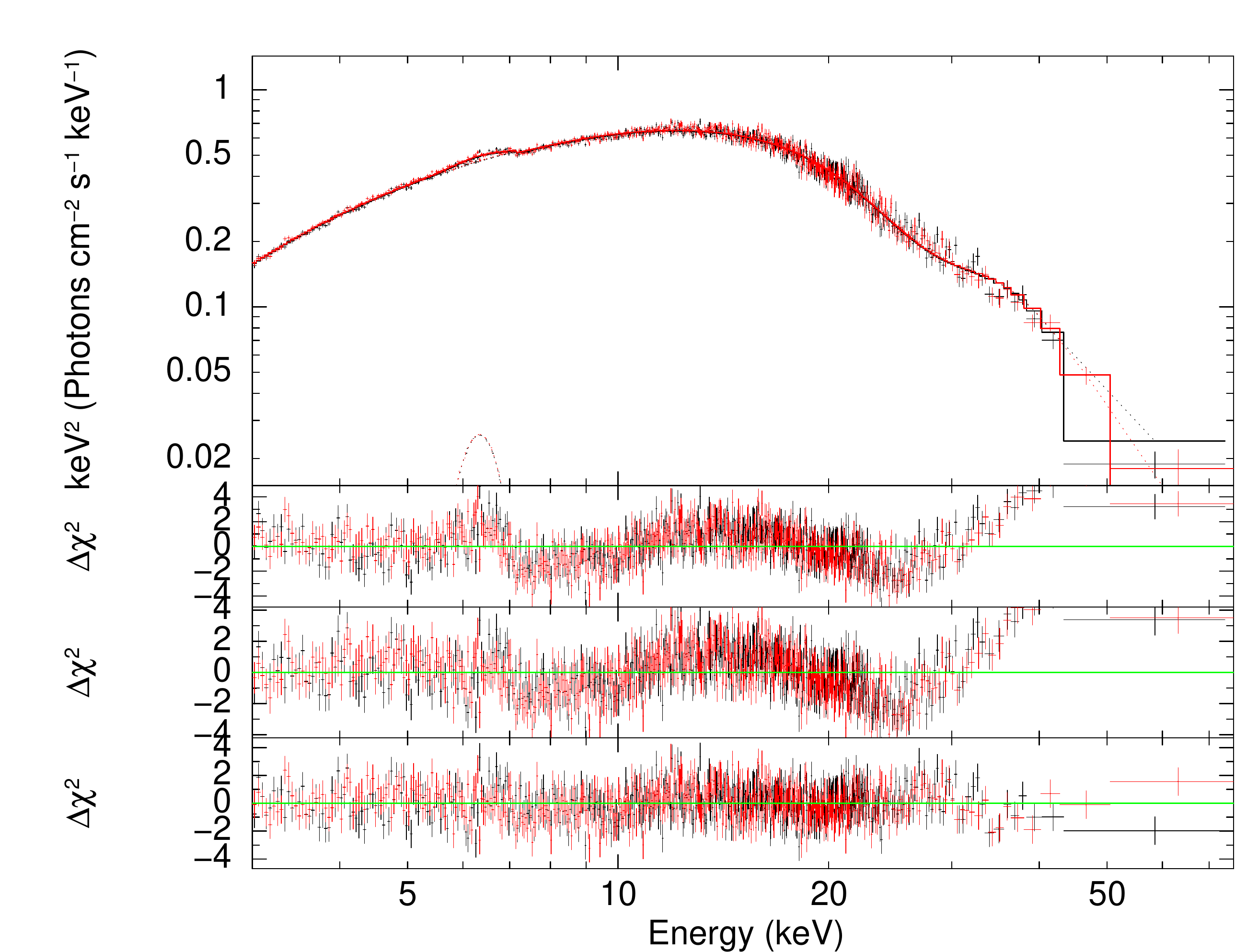}
\end{minipage}
\hspace{0.01\linewidth}
\begin{minipage}{0.33\textwidth}
\includegraphics[height=1.0\columnwidth, angle=-90]{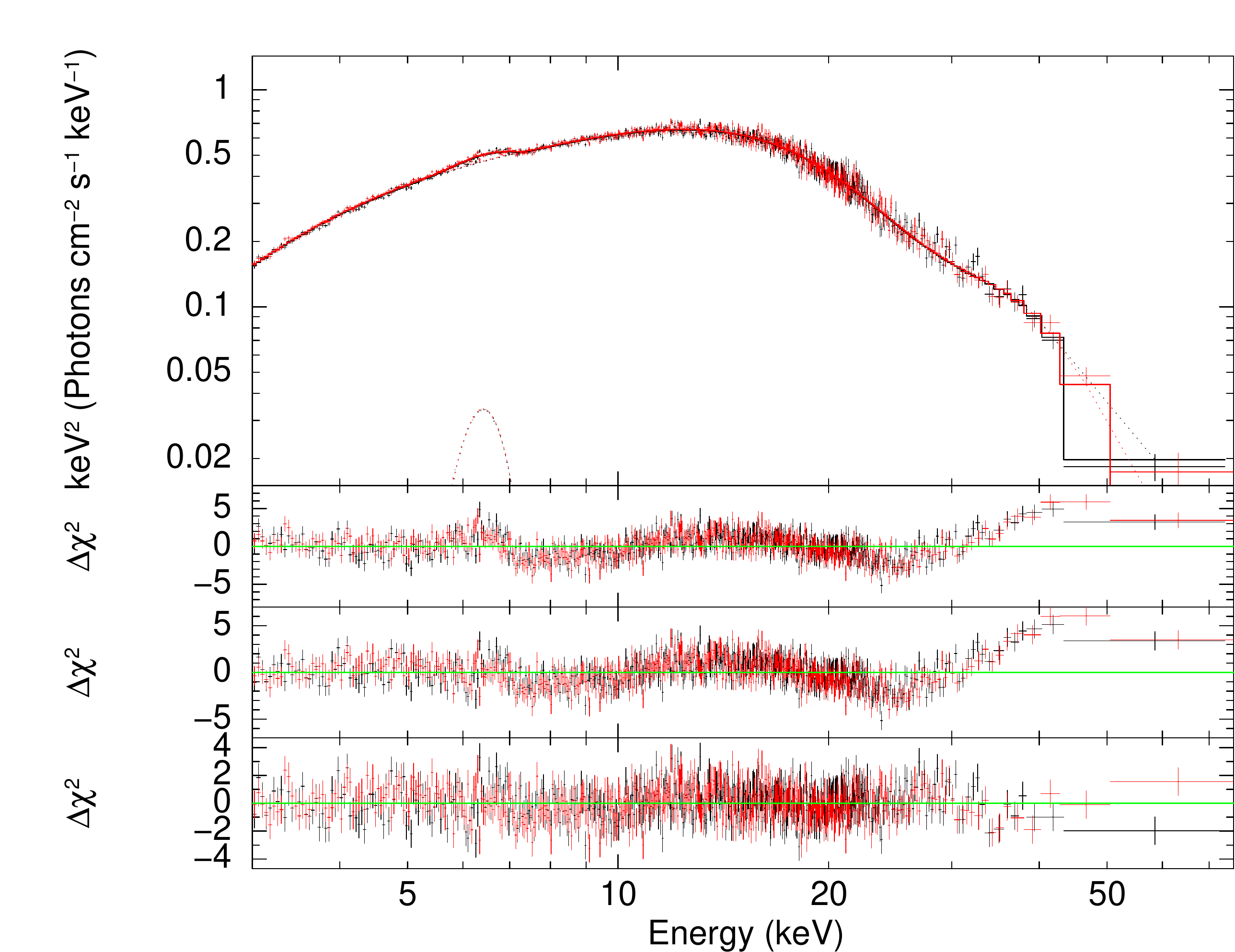}
\end{minipage}
\hspace{0.01\linewidth}
\begin{minipage}{0.33\textwidth}
\includegraphics[height=1.0\columnwidth, angle=-90]{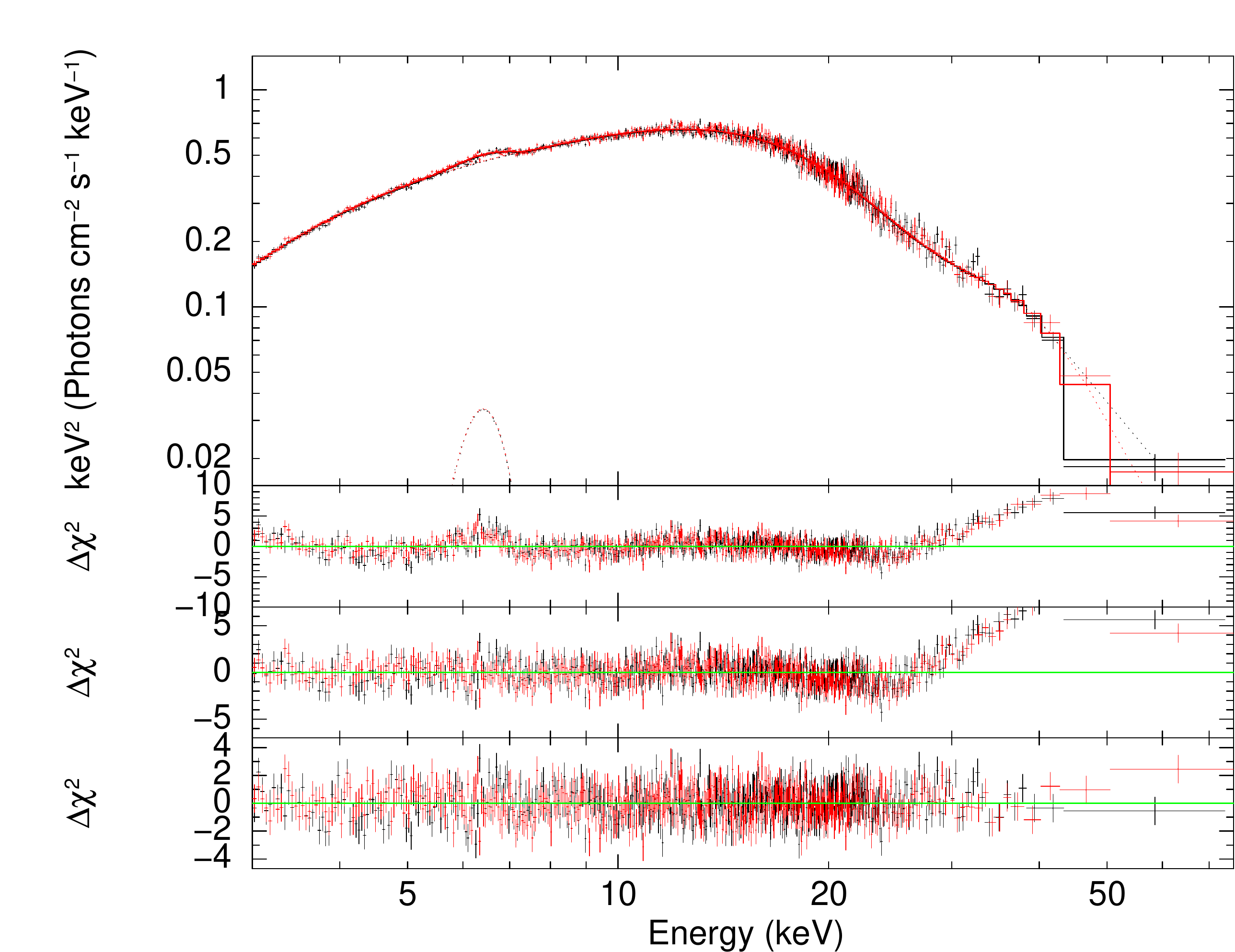}
\end{minipage}
\caption{\textcolor{black}{The folded \emph{NuSTAR} spectra of the 2S 1553-542 along with the residuals for Model I (left), Model II (middle) and Model III (right). For a particular spectra the first (from top) panel represents folded spectra, the second is the residuals left after fitting without GA and GABS model. The third is the residuals without GABS and bottom panel with both GA and GABS.}}
\end{figure*}

 The phase-resolved spectral analysis is helpful in the study of the anisotropic properties of the X-ray emitted by the pulsar around its rotational phase. For this, we divided the pulse period determined in the preceding section into ten equal segments. Using the command XSELECT, Good-time-interval (GTI) files were generated at the determined value of the pulse period. Next, we run NUPRODUCTS for 10 phases using the GTI files created for each phase. After extraction of the data, they were grouped using the GRPPHA tool to obtain a minimum of 20 counts per spectral bin. The creation of phase-resolved spectrum for a periodic source can be efficiently used for comparing the spectrum from different portions of the cycle and hence analyzing the changes that may occur due to rotation of the source. 
 
 The best-fit statistics in the phase-resolved spectroscopy were observed with the implementation of continuum and absorption models - \textsf{(CONSTANT* PHABS* CUTOFFPL)*GABS} in the (3-79) keV energy range. The negative residuals observed in the energy range (24-30) keV due to the cyclotron line was fitted by the absorption model \textsf{GABS}. The results obtained using phase-resolved analysis shows the obvious feature regarding CRSF i.e., pulse-phase variation of the cyclotron line energy. In Figure 3 we have shown the spectrum of the source in the pulse phase 0.0-0.1, in absence of \textsf{GABS} model negative residuals are observed at $\sim$25 keV, with the addition of \textsf{GABS} model the residuals disappears. In absence of the \textsf{GABS} the $\chi^{2}$ is about 1210.91 for 1099 degrees of freedom (dof), after the incorporation of the model the $\chi^{2}$ of the fitting reduces to 1106.75 for 1096 dof. This justifies the addition of \textsf{GABS} model while fitting the spectrum.

\begin{figure}
\includegraphics[scale=0.3, angle=-90]{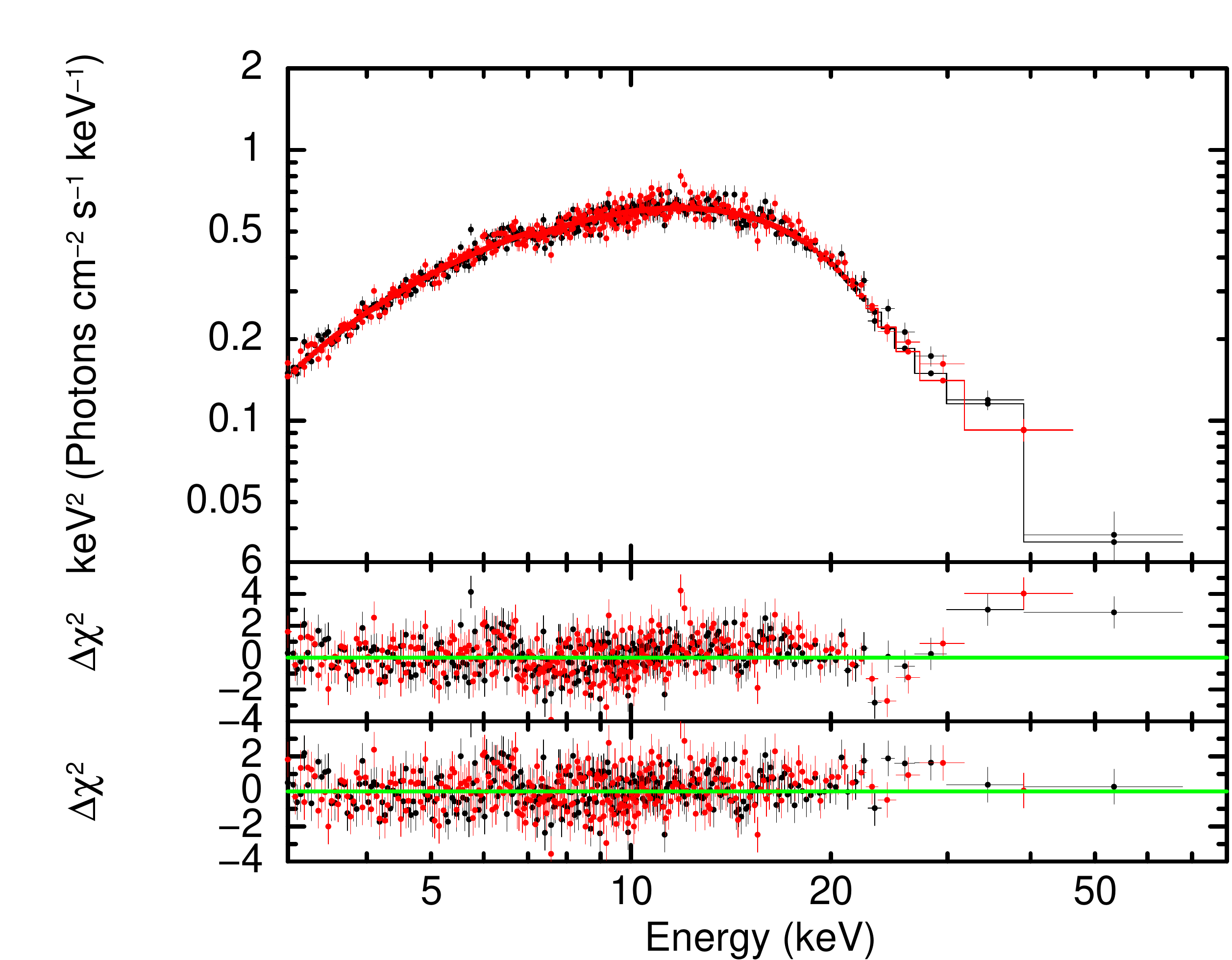}
\caption{The top panel represents the folded spectrum corresponding to one of the phase (0.0-0.1) while the two narrow panels below the folded spectrum represent the residuals without the incorporation of \textsf{GABS} model (middle panel) and after the incorporation of \textsf{GABS} model(bottom panel). Red and black colours represent the \emph{NuSTAR} FPMA and FPMB spectra.}
\end{figure}

\begin{figure}
\includegraphics[scale=0.3]{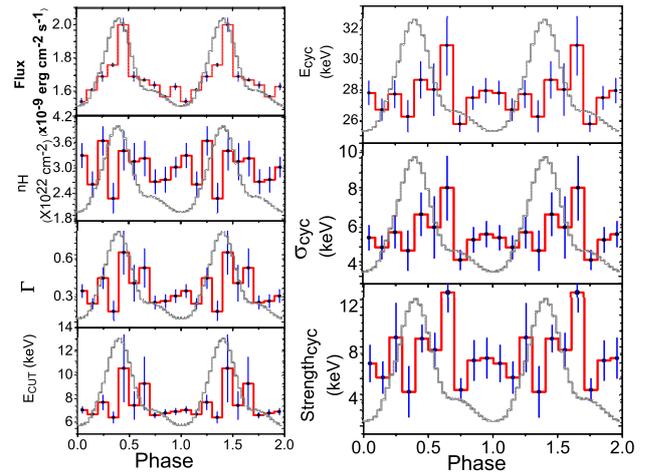}
\caption{\textcolor{black}{Variation of spectral parameters with pulse phase. $\Gamma$ and $E_{CUT}$ represents Photon Index and CUTOFF energy of CUTOFFPL model, $n_{H}$ represents equivalent hydrogen column density for TBABS component, $E_{cyc}$, $\sigma_{cyc}$ and $Strength_{cyc}$ represents the Cyclotron line energy, width and the strength of the GABS component. Flux were calculated within energy range 3-79 keV for the NuSTAR observation.
Errors quoted for each parameter are within 90 \% confidence interval. The continuum pulse profile in the broad energy range 3-79 keV is superimposed in the background in grey colour for visual comparison.}}
\end{figure}

  The variation of different spectral parameters with respect to the pulse phase has been presented in Figure 4. Significant variations in the spectral parameters are observed with the pulse phase. The spectrum is harder (minimum value of $\Gamma$) in between the pulse phase 0.3-0.4 and softer (maximum value of $\Gamma$) in the proceeding phase - 0.4-0.5 when the spectral flux is maximum. In between the phases, 0.7-1.0, where the spectral flux is low the variation of $\Gamma$ is small. Similar variation in $E_{cut}$ can be seen, the minimum of $E_{cut}$  is observed in phase 0.3-0.4 and maximum at 0.4-0.5. The value of $E_{cut}$ is almost constant in between 0.7-1.0. All the parameters of the cyclotron line vary similar manner with pulse-phase. From the pulse-phase variation of the width and the strength of the cyclotron line we observed that they are highly correlated with the centroid energy of the cyclotron line ($E_{cyc}$) and with each other. The Pearson’s correlation coefficient 'r' between $E_{cyc}$ and its width being 0.98 and between $E_{cyc}$ and strength being 0.96. A shift in the phase of the maximum value of the energy of the cyclotron line by 0.2 from the maximum flux is observed.

\subsection{Time resolved spectroscopy of the source using \emph{Swift}-XRT and \emph{NICER} observation of 2021 outburst.}

Time-resolved spectroscopy is often employed for the investigation of fundamental natural processes in real time. Here, we have performed the required time-resolved spectroscopy by considering the different \emph{Swift}-XRT and \emph{NICER} observations. The \emph{Swift}-XRT and \emph{NICER} spectra in 0.7-10 keV energy range have been suitably fitted using PHABS and POWERLAW models. Due to the low count rate of the \emph{SWIFT-XRT} observations, we have implemented C-statistics (Cash 1979) instead of $\chi^{2}$ statistics for obtaining the required spectral fit.
 
The continuum flux of the system in (0.7-10) keV energy range was estimated using the CFLUX model. The variation of the spectral parameters with respect to time has been presented in Figure 5. No significant variation of absorption column density with time is observed, this can be observed from the dash line which represent best fitted line in Figure 5. The absorbed power-law photon index of the pulsar reveals a little increase during the end phase of the outburst. Similarly, it was observed that the flux of the pulsar decreases with time. The correlation coefficient between the photon index and flux is -0.55. Thus with the decrease in the flux, there is an increase in the photon index.

 Similar trends in the photon-index and flux were observed using the \emph{NICER}. From Figure 5 we can see that the photon index increases with a decrease in the flux, the correlation coefficient between photon index and flux in this case being -0.86. Considering the values of photon-index from both \emph{Swift-XRT} and \emph{NICER} simultaneously we got the correlation to be equal to -0.4 between photon-index and flux. Thus there is a mild softening of the spectrum at the end of the outburst.

\begin{figure}
\centering
\includegraphics[scale=0.32]{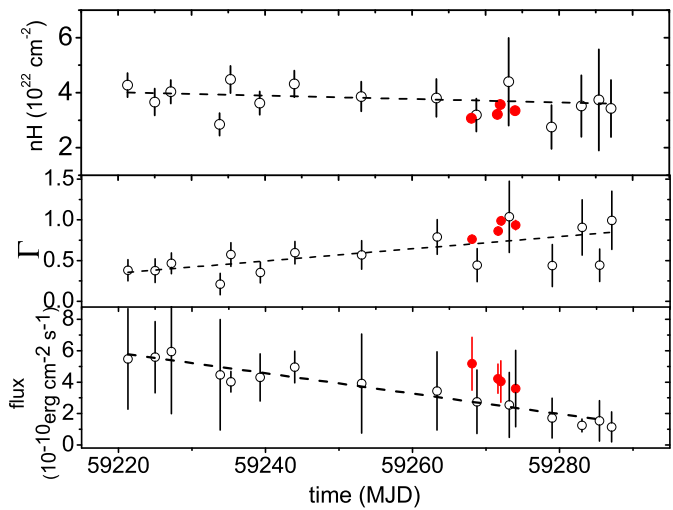}
\caption{Variation of column density ($top$) photon index ($middle$) and flux ($bottom$) of the pulsar with time during 2021 outburst. The hollow circle is for \textit{Swift}-XRT and the filled red circle is for \textit{NICER}. The flux shown is in 0.7-10 keV energy range. The dash line indicates best fitted straight line.}
\end{figure}

\subsection{\textbf{Time resolved spectroscopy of the pulsar using \emph{RXTE}/PCA observation of the 2008 outburst.}}

\subsubsection{\textbf{Colour-colour diagram - }}

In order to study whether the long-term variation of the spectral properties of the pulsar depends on the properties of the light curves of the pulsar, we plotted hardness intensity diagram (HID) during the 2008 outburst. The hardness ratio is defined as the ratio of background subtracted light curves in 7-15 keV and 3-7 keV energy ranges. The HID diagram of the pulsar is shown in Figure 6. Two different branches which are observed in the HID - horizontal branch (HB) and two diagonal branch (DB). The names HB and DB are as defined by Reig \& Nespoli (2013). 

\begin{figure}
\centering
\includegraphics[scale=0.35]{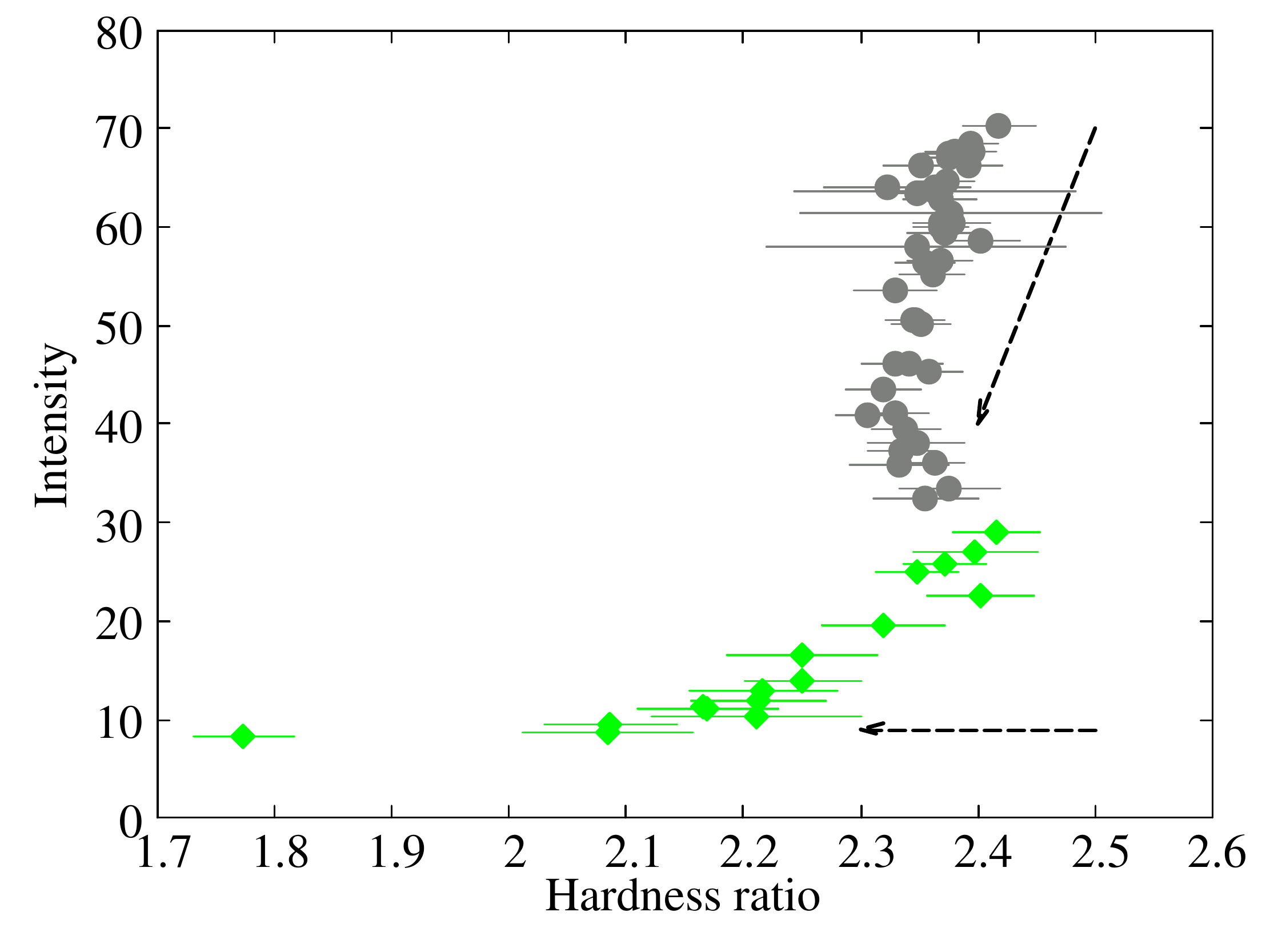}
\caption{\textit{Upper panel} -Hardness-intensity diagram of the pulsar. HB (green) stands for horizontal branch and DB (grey) for the diagonal branch.}
\end{figure}

\begin{table*}
\centering
\begin{tabular}{c c c c c}
\hline
\hline
$Variables$  &$\Gamma-F$ &$E_{cut}-F$ & $HC-F$ & $SC-F$ \\
 & $HB/DB$ & $HB/DB$ & $HB/DB$ & $HB/DB$ \\
\hline
$r$ &-0.57/0.69 & -0.24/0.38  &-0.53/0.09 & 0.77/0.61 \\
$CI (\%)$ &90/99 & 80/98 & 99/85 & 98/99 \\
\hline

\end{tabular}
\caption{Table showing the correlation between different parameters. Here $r$ indicates the values of Pearson's correlation co-efficient and $CI$ represents the confidence intervals in percentage and are greater than the values given in the table. $F$ stands for $\emph{RXTE}$/PCA flux in 3-23 keV energy range.}
\end{table*}

\begin{figure}
\centering
\includegraphics[height=8 cm, width= 9 cm, angle= -90]{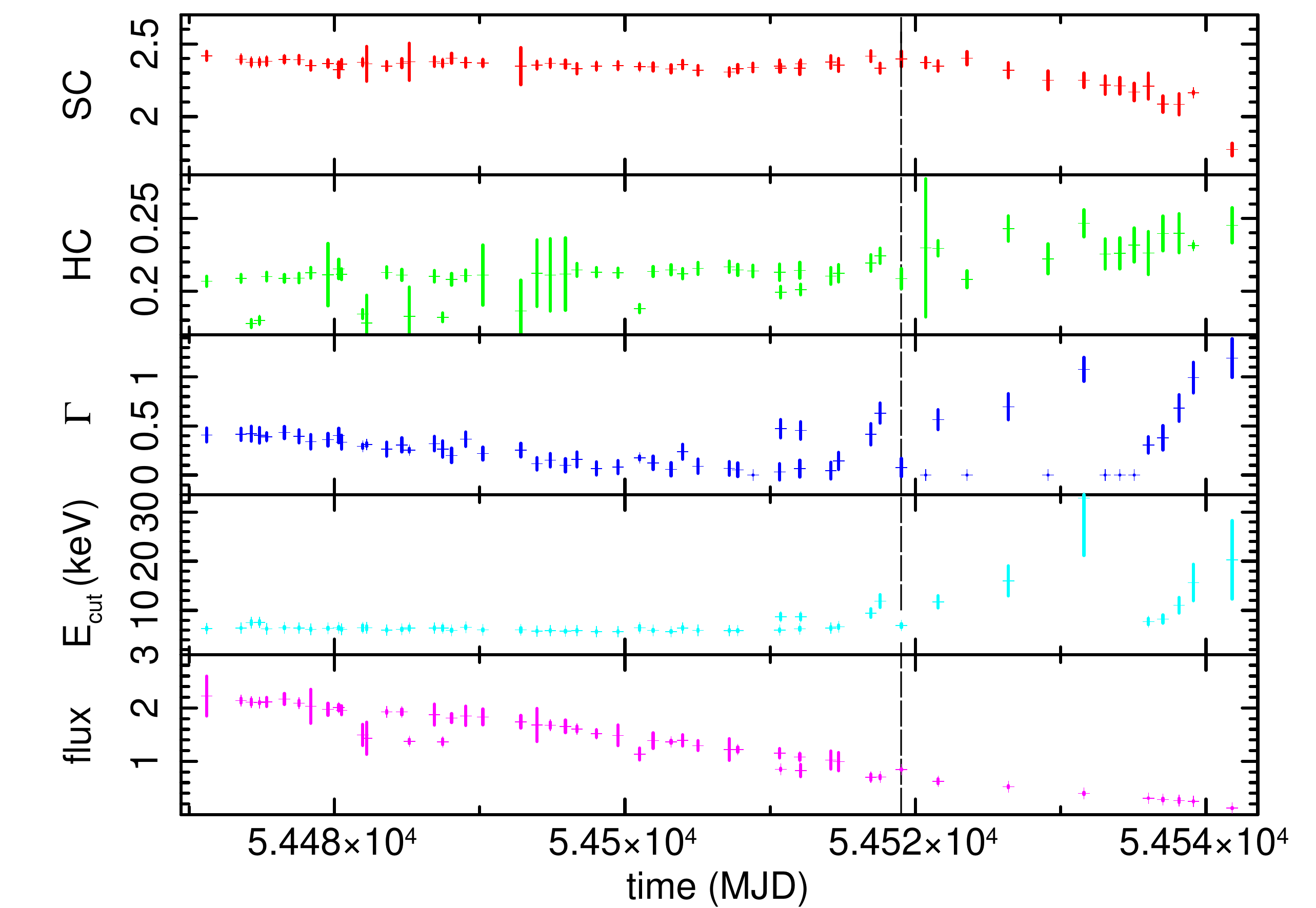}
\caption{Variation of soft colour (SC), hard colour (HC), photon-index ($\Gamma$), cutoff energy ($E_{c}$) and flux. The flux is in the scale of $10^{-9}$ erg cm$^{-2}$ s$^{-1}$. The vertical dash line indicates transition from diagonal to horizontal branch.}
\end{figure}
\begin{figure}
\centering
\includegraphics[scale=0.30]{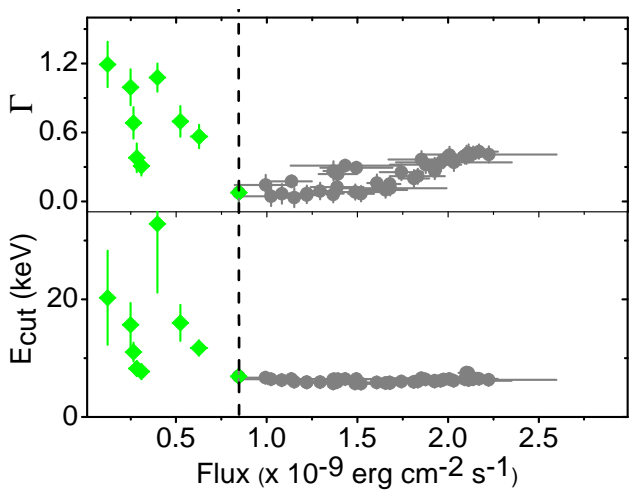}
\caption{Variation of photon index (top panel) and cut-off energy (bottom panel) with flux.}
\end{figure}

\subsubsection{\textbf{Spectral variability - }}

 The correlation of the colours to time has been observed and presented in this section. We have also tried to explore the variations of spectral parameters and the continuum flux of the system with time by considering \emph{RXTE}/PCA observations. The \emph{RXTE}/PCA spectra in 3-23 keV energy range were well fitted using PHABS and CUTOFFPL models. Initially, for the first few observations, the emission feature signature in the spectra was modeled by adding the GAUSSIAN component during the fitting procedure. However, for other observations, it was exceedingly difficult to generate optimum values of the parameters corresponding to the GAUSSIAN model. Hence, we were bound to drop this model for generating the best fit statistics for the system. We have suitably fitted the spectra corresponding to all the observations without incorporating the GAUSSIAN model, the average value of $\chi_{\nu}^{2}$ of the fitting is $\sim$ 1.24 with an average of 45 dof. The flux of the system in the specified energy range was estimated using the model CFLUX. The variation of spectral parameters corresponding to the CUTOFFPL model, along with the variations of hard and soft colours with time has been shown in Figure 7. The vertical line in Figure 7 corresponds to the point of transition from the diagonal to the horizontal portion of HID. The \emph{RXTE/PCA} flux in 3-23 keV energy range is $\sim$(8.46$\pm$0.13)$\times$10$^{-10}$ erg cm$^{-2}$ s$^{-1}$, the corresponding extrapolated bolometric flux in 0.001-100 keV energy range is $\sim$(1.02$\pm$0.05)$\times$10$^{-9}$ erg cm$^{-2}$ s$^{-1}$. The distance to the source is about $\sim$ 20 kpc (Tsygankov et al. 2016) which means that the bolometric luminosity is about (4.88$\pm$0.24)$\times$10$^{37}$ erg s$^{-1}$. Figure 8 shows variation of photon index and $E_{cut}$ with the flux. The correlation of photon index ($\Gamma$), cutoff energy ($E_{cut}$ ), soft (SC), and hard (HC) colours with flux for different branches have been presented in Figure 9. Interestingly, negative residuals revealing absorption characteristics were prominently observed in the spectra in the 8-12 keV energy range. The residuals at 8-12 keV have been reported as the 10 keV feature by Coburn et al. (2002). As suggested by Reig \& Nespoli (2013), 10 keV feature can be fitted by incorporating an absorption model. The inclusion of an absorption model improved the spectral fit (eg. in case of the spectrum corresponding to the obsID 93426-01-04-02 the $\chi_{\nu}^{2}$ decreases from 1.44 with 42 dof to 1.27 with 39 dof after an inclusion of absorption model). Though, the absorption feature was suitably modeled using an absorption model, it would be inappropriate to quote it as a CRSF line as this feature was not observed in the pulsar spectra of other known missions. The origin of this feature is not known yet, however, Coburn et al. (2002) suggested that it might be due to the shortcoming of the simple phenomenological models used in fitting the spectrum.

The correlation strength amongst different parameters can be estimated in terms of Pearson’s correlation coefficient 'r'. The significance of the correlations is obtained by considering a probabilistic interpretation (Reig \& Nespoli 2013). The chance probability p can be estimated by considering Student’s two tailed $t$-distribution, such that $p=T(N-2,t)$ where $t$ is expressed as:

\begin{eqnarray*}
t=r\sqrt{\dfrac{N-2}{1-r^2}}
\end{eqnarray*}

In the above expression, N represents the number of data points and N-2 signifies the degrees
of freedom of the system. A lower value of p expresses a lower possibility of the existence of correlation by chance and vice-versa. For $p<\alpha$, the existence of the correlation by chance becomes significantly low where the factor $(1-\alpha)\times100$ gives an estimate about the confidence interval. As evident from Figure 9, the power-law photon index ($\Gamma$) in the HB is found to possess a negative correlation with respect to the flux while it bears a positive correlation in the DB ($r=0.69,99\%$). The correlation coefficient between $E_{cut}$ and flux for HB and DB are -0.24 (80\%) and 0.38 (98\%) respectively. So weak anti-correlation between $E_{cut}$ and flux exists for HB branches of HID and weak positive correlation exist between the two quantities for DB of HID. The existence of two branches (DB and HB) is a noticeable feature that reveals a concrete assertion about the transition in the accretion schemes of the pulsar.
 
 Furthermore, the correlation between flux and the two colours has also been analyzed. Hard colours are observed to possess an anti-correlation with respect to the continuum flux in HB with the relevant Pearson’s correlation coefficient (r) estimated to be -0.53 (99\%). No correlation between the two quantities exist in case of the DB. It is obvious that the soft colour bears a positive correlation with the continuum flux in both HB $( r=0.77, 98\%)$ and DB  $(r=0.61, 99\%)$. The power-law photon index $\Gamma$ is found to diminish with the passage of time along the DB which is in accordance with its positive correlation with respect to the flux. Along the HB, the photon index follows an increasing trend with time which is consistent with our observation regarding its anti-correlation with the flux near the end of the outburst.
\begin{figure*}
\begin{minipage}{0.3\textwidth}
\includegraphics[scale=0.3]{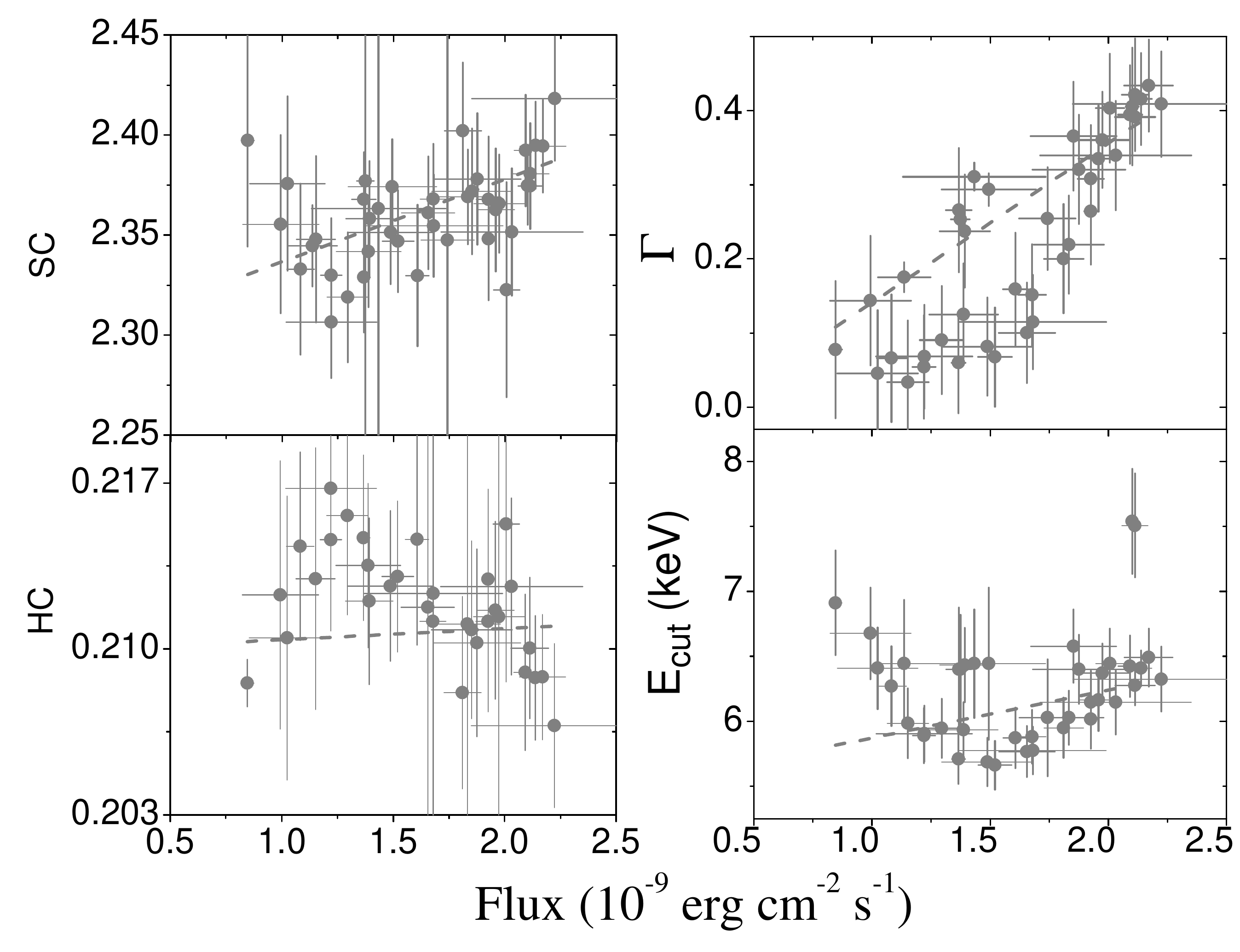}
\end{minipage}
\hspace{0.2\linewidth}
\begin{minipage}{0.3\textwidth}
\includegraphics[scale=0.3]{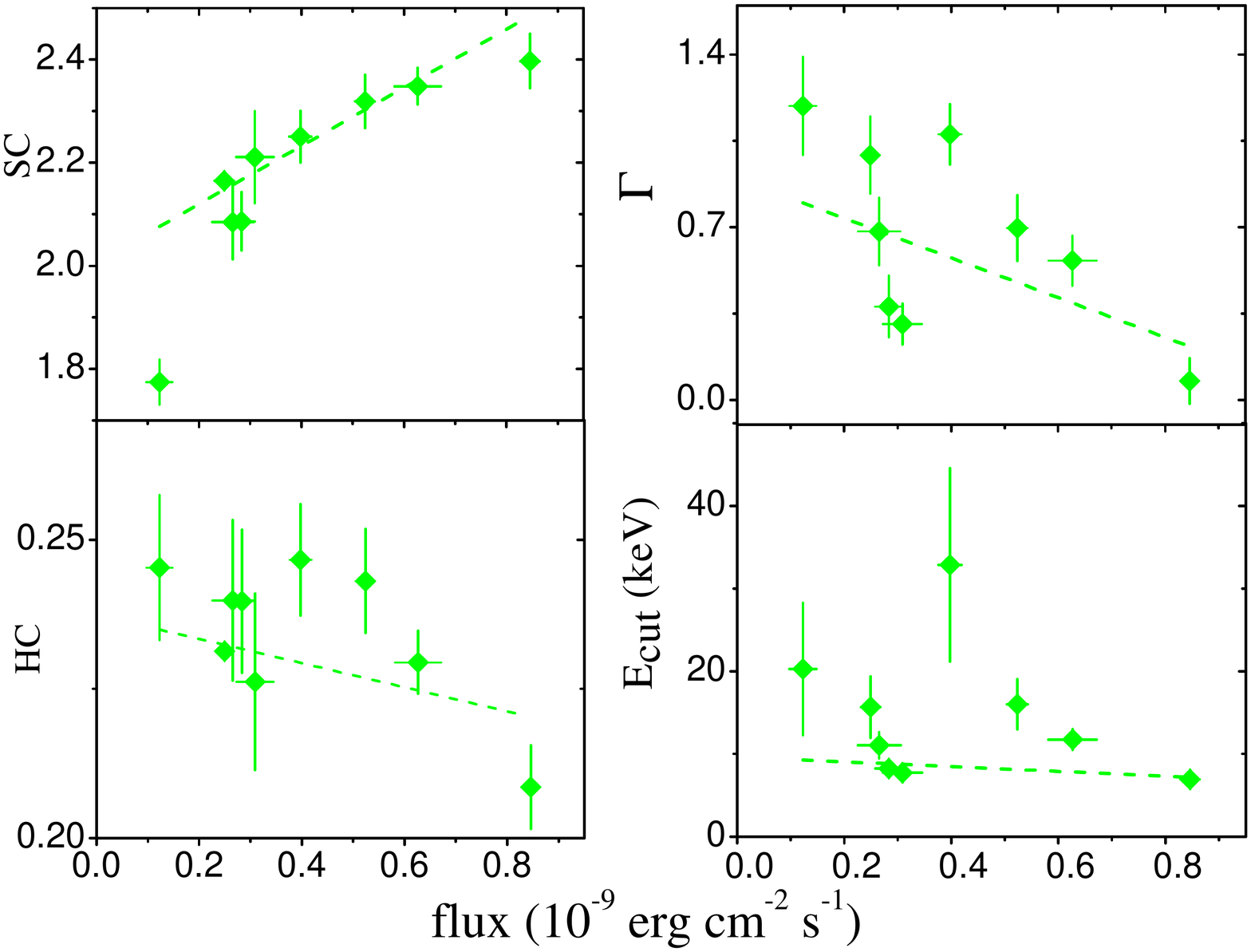}
\end{minipage}
\caption{\textit{Left} - Variation of Soft colour, Hard colour, photon index ($\Gamma$) and cutoff energy ($E_{cut}$) with flux for diagonal branches of HID. The dashed grey lines are best fitted straight line for DB. \textit{Right} - Variation of spectral parameters and colours with flux for HB. Dashed green lines are best fitted straight line.}
\end{figure*}

\section{Spin-up rate and flux}

  During a major outburst when the external accreting torque acts along the same direction in which a neutron star rotates then the spinning up of the pulsar is observed (Ghosh \& Lamb 1979) and it is direct indication of the presence of accretion disc around the neutron star. During type II outbursts of Be/X-ray binaries it is proposed that  the accretion disc is formed (Motch et.al 1991 ; Wilson  et al 2008) and spinning up of a pulsar can be due to disc accretion as the direct accretion will not produce spinning up (Reig 2011). \textcolor{black}{In order to understand the spin evolution of the source 2S 1553-542, we made use of source spin frequency ($\nu$) history provided by FERMI GBM Accreting Pulsars Program (GAPP) team during the recent outburst in 2021 (Figure 9).} The decade long observations of X-ray pulsars using FERMI GBM is reported by Malacaria et al. (2020). The spin frequency provided by GAPP is orbital corrected. The estimation of the required spin-up rate of the system has been carried out by a linear fitting of three consecutive frequencies (Serim et al. 2022). The slope of the plot represents the rate of change of spin frequency $(\dot{\nu})$ and its corresponding uncertainty represents the 1-sigma uncertainty associated with the measurement of the spin-up rate. Here, we have made use of the flux corresponding to the \emph{Swift}-XRT observations to analyze the variation of the spin-up rate of the system. The flux in 0.7-10 keV energy range was extrapolated to bolometric flux in the energy range 0.001-100 keV. The observed variation has been furnished alongside in Figure 10. Since the flux corresponding to the time where the spin-up rate is measured is not explicitly known, so we have made use of the linear interpolation technique on the time evolving \emph{Swift}-XRT flux to estimate the required flux of the system at that point in time.

\begin{figure*}
\begin{minipage}{0.3\textwidth}
\includegraphics[height=1.1\columnwidth, angle=0]{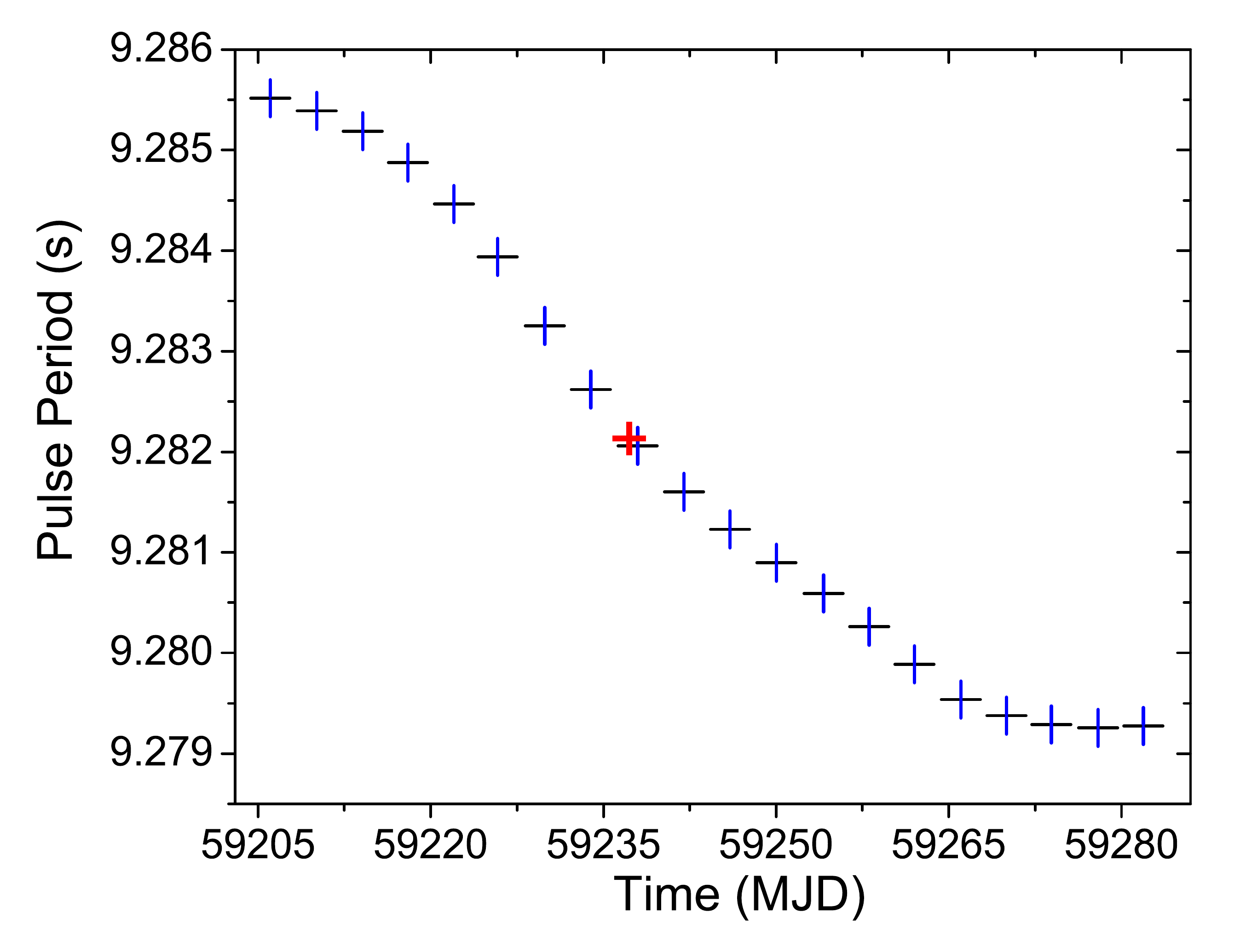}
\end{minipage}
\hspace{0.2\linewidth}
\begin{minipage}{0.3\textwidth}
\includegraphics[height=1.1\columnwidth, angle=0]{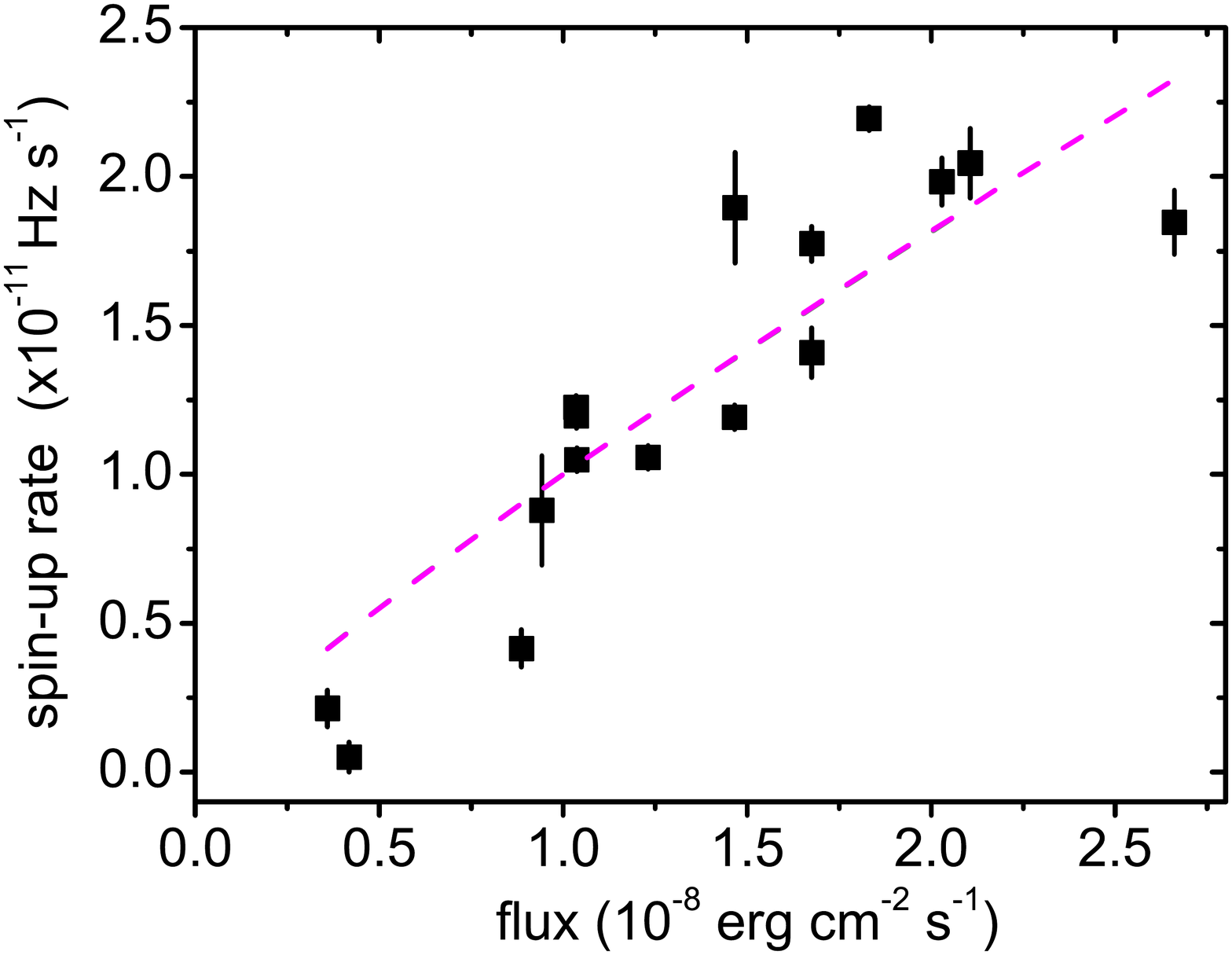}
\end{minipage}
\caption{The left panel represents the spin evolution of the source using Fermi-GBM data, the period corresponding to the \emph{NuSTAR} observation is shown by red ``+'' sign. The right panel represents the variation of the spin up rate with the flux. The dash lines in magenta indicates the best fitted line.}

\end{figure*} 

To understand the variation of $\dot{\upsilon}$ with respect to flux, the fitting has been carried out by implementing a power-law model \textcolor{black}{($\dot{\upsilon} \propto F^{a}$)}, as shown by the dashed magenta line in Figure 10. The best-fitted value of the power-law index $a$ is $\sim$0.862$\pm$0.080. In accretion torque model the spin-up rate of the system is found to vary with the mass accretion rate ($\dot{M}$) as  \textcolor{black}{ $\dot{\upsilon}\;\propto\; \dot{M}^{6/7}$ }(Ghosh \& Lamb 1979). It is obvious that the net flux of the system is directly dependent on the mass accretion rate, thus the spin-up rate and the flux of the system bear a relationship with one another similar to the one specified above. The aforementioned estimated \textcolor{black}{power-law index} (0.862) from our analysis is close to the theoretical exponent of 6/7 ($\sim$0.857), which implies our result is in accordance with the theoretical predictions. Hence, our estimation and analysis can be considered to be reliable as it reveals a very close resemblance with the \textcolor{black}{model proposed by Ghosh\& Lamb 1979}.

\section{Discussion}
\subsection{Pulse-phase dependent of cyclotron line}

\textcolor{black}{The cyclotron line parameters are} dependent on the viewing angle i.e. the pulse phase of the system (Isenberg et al. 1998; Araya-Gòchez \& Harding 2000; Mukherjee \& Bhattacharya 2012). The numerical simulations of the cyclotron line \textcolor{black}{suggest} that the line parameters are found to vary by about 10-20\% with the pulse phase (Araya-Gòchez \& Harding 2000; Sch{\"o}nherr et al. 2007; Mukherjee \& Bhattacharya 2012).\textcolor{black}{The CRSF absorption feature was not observed in the \emph{RXTE} data from 2008, but is prominently observed in 2021}. These parameters can vary by more than 20\% up to 40\% \textcolor{black}{(Vasco et al. 2013)} over the pulse phases, for example the variation of centroid energy in case of Her X-1 (first detected CRSF) is of the order of 25\% (Staubert et al. 2014). The first variations of the cyclotron line with respect to the pulse phase of the system were reported by Voges et al. (1982); Soong et al. (1990). The variations of the CRSF energy are a consequence of sampling different heights of the line forming region as a function of pulse phase \textcolor{black}{(Staubert et al. 2014).} In our study, the centroid energy of the cyclotron line was found to vary with the pulse phase. The evidence from our observations is revealing clear signatures of the usual behavior of CRSFs which includes an effective \textcolor{black}{variability} of the CRSF energy with pulse phase. Moreover, the line energy can vary with pulse phase, luminosity, time, and phase of the super-orbital period (Staubert et al. 2014). In this work, it was observed to vary between 25.81-30.90 keV with pulse phase and which corresponds to almost $11\%$ about the mean. \textcolor{black}{The other parameters of the cyclotron line (i.e., width and \textcolor{black}{strength}) were also found to exhibit \textcolor{black}{variation} with respect to the pulse phase.} The specific variation of the parameters is presented alongside in \textcolor{black}{Figure 4.} The centroid energy, width, and strength of the cyclotron line were found to possess identical variations as evident from the \textcolor{black}{Figure 4.} The width and the strength of the cyclotron line is found to vary by more than 20 \% about their mean value. They also bear strong positive correlation with the centroid energy of the cyclotron line which are general \textcolor{black}{features} of cyclotron line.

\textcolor{black}{We have observed a shift in the phase of the maximum of the cyclotron line by 0.2 of that of the pulse profile. The shift in the maximum of the pulse profile and the cyclotron line can be explained using the reflection model of the cyclotron line proposed by Poutanen et al. (2013). According to the model the formation of the cyclotron line occurs when the radiation emitted by the accretion column is reflected from the surface of the neutron star. So the reflected spectrum consist of the cyclotron line  whereas the total flux at any particular pulse phase consist of flux  directly from the accretion column and reflected part. The observed variation of the cyclotron line with the pulse phase can be explained on the basis of contribution of the two components. The ratio of the reflected to the direct emission depends on the total fraction of emission intercept and on the beaming pattern of the two components. The reflected part is expected to exhibit broad pencil-like beam pattern, which is direct along the magnetic field lines. However the beam pattern of the direct component is complicated and depends on the luminosity of the pulsar. Since the cyclotron line is only present in the reflected component we may expect it to be anti-phase with the direct emission. Thus maximum of the pulse profile shifts with respect to the maximum of the cyclotron line energy profile (Lutovinov et al. 2015). Lutovinov et al. (2015) explained the observed phase-shift of $\sim$0.2-0.35 between the maximum of the pulse profile and the cyclotron line in the low luminosity state (of the order of 10$^{37}$ erg s$^{-1}$) of V 0332+53 using the reflection model.}

\subsection{HID diagram and the Critical luminosity}

The HID diagram of the pulsar corresponding to the 2008 Type II outburst bears interesting characteristics which add more clarity regarding the transition of the accretion schemes \textcolor{black}{(Reig \& Nespoli 2013)}. Two diagonal branches along with a noticeable horizontal branch is observed in HID. At higher or intermediate fluxes, the majority of the observations lie along the \textcolor{black}{diagonal branch}. However, during the decay phase of the outburst the flux of the system is found to diminish and source moves along the horizontal branch. Furthermore, the spectral properties of the source were also relevantly observed to exhibit corresponding variations. The power-law photon index ($\Gamma$) along the HB is found to possess a negative correlation with respect to the flux while it bears a positive correlation along the DB. The correlation between flux and the two colours has been observed and it is found that the hard colours possess an overall anti-correlation to the continuum flux of the system in HB. It is seen that the soft colour bears a positive correlation with in both HB and DB. After considering the distinctly different evolution of the parameters in the two branches DB and HB as shown by the Figure 9, it would be conclusively evident to express the fact that the system has undergone a transition in the accretion scheme (Reig \& Nespoli 2013). This is relevant from the HID (Figure 6) where the distinct transition from DB to HB are prominently visible. The existence of two spectral states of the pulsar can be as a consequence of two different accretion regimes (Reig \& Nespoli 2013). The photon index follows a decreasing trend with flux in the sub-critical regime while it exhibits an increasing trend in the supercritical regime. Hence, the diagonal and horizontal branches reveal the accretion regimes of the pulsar. The transition in the accretion regime of the source is  characterized by a critical luminosity ($L_{crit}$, Becker et al. 2012).
   \begin{equation}
    L_{critical}=1.5\;\times\;10^{37}\;\times\;B^{15/16}_{12} erg \;s^{-1}
    \end{equation}
where $B_{12}$ is the surface magnetic field in units of $10^{12} G$. From the cyclotron line energy, we can constrain the magnetic field of the neutron star using equation 1. \textcolor{black}{For cyclotron line of energy $\sim$ 27 keV the estimated magnetic field using equation 1 is $\sim$ 2.9$\times$10$^{12}$ G}. The value of $L_{crit}$ has been reported by Malacaria et al. (2022) to be $\sim$ 4$\times$10$^{37}$ erg s$^{-1}$. In this work, the luminosity corresponding to the transition between the two branches is about $(4.88\pm0.24)\;\times\;10^{37}erg\;s^{-1}$ which is close to the value of the critical luminosity obtained by Malacaria et al. (2022). Thus the existence of two different branches hints at the two different accretion regimes - subcritical and supercritical regimes - separated by the critical luminosity.
    
    In the subcritical accretion regime, an increase in luminosity causes the height of the emission zone in the accretion column to approach the surface of the neutron star \textcolor{black}{(Becker et al. 2012)}. Similarly, in the supercritical accretion state, the emission zone is shifted up in the accretion column. The Comptonization of photons occurs in the region (sinking zone) between the radiative shock and the neutron star (Becker \& Wolff 2005a, 2005b, 2007). The dimensions corresponding to this region are fairly small during supercritical state while its height increases with increasing luminosity (please refer to eq. 40 of Becker et al. 2012). The region between the radiative shock and the neutron star is characterized by a balanced rate of diffusion and advection which reduces the velocity of the comptonising electrons by a significant extent. Consequently, the bulk Comptonization cannot provide enough energy to the photons to excite them to higher energy levels leading to the softness of the spectrum with increasing luminosity. This explains the positive correlation of the photon index with flux along the DB. Now, switching our discussion to the sub-critical state, the height of the emission region is known to be a few kilometers but is found to diminish gradually with increasing luminosity (see eq. 51 of Becker et al. 2012). The decrease in the size of the sinking region with an increase in luminosity leads to an increase in the optical depth thereby increasing the hardness of photons (Reig \& Nespoli 2013). This is in agreement with the hardening (decrease in $\Gamma$) of the spectrum with an increase in flux or softening of the spectrum with a decrease in flux along the HB. This may also be inferred to be the reason behind the spectral softening of \emph{Swift}-XRT spectra towards the end of the recent outburst.

 \section*{Acknowledgements}
 The authors of the paper would like to thank the anonymous reviewers for their vital suggestions to improve the quality of the paper. This research has made use of publicly available data of pulsar provided by the NASA HEASARC data archive. The \textit{FERMI} data used in the research is provided by \textit{FERMI}-GAPP team. 
 
 \section*{Data availability}
 
 The observational data used in the research can be accessed from the HEASARC data archive.












\begin{theunbibliography}{}
\vspace{-1.5em}
\bibitem{latexcompanion}
Araya-G{\`o}chez, R. A., \& Harding, A. K. 2000, ApJ, 544, 1067
\bibitem{latexcompanion}
Arnaud, K. A. 1996, in Astronomical Data Analysis Software and Systems V,
eds. G. H. Jacoby, \& J. Barnes, ASP Conf. Ser., 101, 17
\bibitem{latexcompanion}
Basko, M. M., Sunyaev. R. A., 1976, MNRAS, 175, 395
\bibitem{latexcompanion}
Becker, P. A., \& Wolff, M. T. 2005a, ApJ, 621, L45
\bibitem{latexcompanion}
Becker, P. A., \& Wolff, M. T. 2005b, ApJ, 630, 465
\bibitem{latexcompanion}
Becker, P. A., \& Wolff, M. T. 2007, ApJ, 654, 435
\bibitem{latexcompanion}
Becker, P. A., Klochkov, D., Sch{\"o}nherr, G., et al. 2012, A\& A, 544, A123
\bibitem{latexcompanion}
 Belloni, T. 2010, in The Jet Paradigm – From Microquasars to Quasars (Berlin
Heidelberg: Springer-Verlag) Lect. Notes Phys., 794, 5
\bibitem{latexcompanion}
Boldin, P. A., Tsygankov, S. S., \& Lutovinov, A. A. 2013, Astronomy Letters, 39, 375
\bibitem{latexcompanion}
Bildsten L. et al., 1997, ApJS, 113, 367
\bibitem{latexcompanion}
Burnard D. J., Arons J., Klein R. I., 1991, ApJ, 367, 575


\bibitem{latexcompanion}
Caballero I., Wilms J., 2012, Mem. Soc. Astron. Italiana, 83, 230
\bibitem{latexcompanion}
Cash, W. 1979, ApJ, 228, 939
\bibitem{latexcompanion}
Coburn, W., Heindl, W. A., Rothschild, R. E., Gruber, D. E., Kreykenbohm, I., Wilms, J., Kretschmar, P. \& Staubert, R 2002, 580, 394 
\bibitem{latexcompanion}
dal Fiume D. et al., 2000, Adv. Sp. Res., 25, 399

\bibitem{latexcompanion}
Davidson, K. \& Ostriker, J. P. 1973, ApJ, 179, 58

\bibitem{latexcompanion}
Deeter J. E., Boynton P. E., Pravdo S. H., 1981, ApJ, 247, 1003
\bibitem{latexcompanion}
Filippova E. V., Tsygankov S. S., Lutovinov A. A., Sunyaev R. A., 2005,  Astronomy Letters, 31, 729
\bibitem{latexcompanion}
Gehrels, N., Chincarini, G., Giommi, P., et al. 2004, The
ApJ, 611, 1005
\bibitem{latexcompanion}
Gendreau, K., \& Arzoumanian, Z. 2017, Nature Astronomy, 1, 895
\bibitem{latexcompanion}
Ghosh P., Lamb F. K., 1979, ApJ , 234, 296

\bibitem{latexcompanion}
Harding, A. K. 1994, in American Institute of Physics Conference Series, Vol.308, The Evolution of X-ray Binariese, ed. S. Holt \& C. S. Day, 429

\bibitem{latexcompanion}
Hasinger, G., \& van der Klis, M. 1989, A \& A, 225, 79
\bibitem{latexcompanion}
Homan, J., \& Belloni, T. 2005, Ap\& SS, 300, 107
\bibitem{latexcompanion}
Isenberg, M., Lamb, D. Q., \& Wang, J. C. L. 1998, ApJ, 505, 688
\bibitem{latexcompanion}
Jahoda, K., Swank, J. H., Stark, M. J., et al. 1996, EUV, X-ray and Gamma-ray
Instrumentation for Space Astronomy VII, eds. O. H. W. Siegmund, \& M. A.
Gummin, SPIE 2808, 59
\bibitem{latexcompanion}
Jenke P., Colleen W.-H., Malacaria C., ATel, 14301
\bibitem{latexcompanion}
Kelley R. L., Ayasli S., Rappaport S., 1982, IAUC, 3667, 3
\bibitem{latexcompanion}
Krimm H. A. et al., 2007, ATel, 1345, 1



\bibitem{latexcompanion}
Lutovinov A. A., Tsygankov S. S., 2009,  Astronomy Letters, 35, 433
\bibitem{latexcompanion}
Lutovinov S. S., Tsygankov S. S., Suleimanov V. F., Mushtukov A. A., Doroshenkov V., Nagirner D. I., Poutanen J., 2015, MNRAS, 448, 2175
\bibitem{latexcompanion}
Lutovinov A. A., Buckley, D. A. H., Townsend, L. J., Tsygankov, S. S., Kennea J., 2016, MNRAS, 462, 3823

\bibitem{latexcompanion}
Lyubarskii, Y. E. \& Syunyaev, R. A. 1982, Soviet Astronomy Letters, 8, 330
\bibitem{latexcompanion}
 Malacaria C., Jenke P., Roberts O. J., Wilson-Hodge C. A., Cleveland W.H. and Mailyan B., 2020, ApJ, 896, 90 
\bibitem{latexcompanion}
 Malacaria C. et al. 2021, ATel, 14348 
\bibitem{latexcompanion}
 Malacaria C. et al. 2022, ApJ, 927, 194
\bibitem{latexcompanion}
Meegan C. et al, 2009, 702,791 
\bibitem{latexcompanion}
Motch C., Stella L., Janot-Pacheco E., Mouchet M., 1991, ApJ, 369, 490
\bibitem{latexcompanion}
Mukherjee D., Bhattacharya D., 2012, MNRAS, 420, 720
\bibitem{latexcompanion}
Nelson R., Salpeter, E., \& Wassermann, I. 1993, ApJ, 418, 874
\bibitem{latexcompanion}
Pahari M. \& Pal S., 2012, MNRAS, 423, 3352
\bibitem{latexcompanion}
Poutanen J., Mushtukov A. A., Suleimanov V. F., Tsygankov S. S., Nagirner
D. I., Doroshenko V., Lutovinov A. A., 2013, ApJ, 777, 115
\bibitem{latexcompanion}
Reig P., 2008, A \& A, 489, 725R
\bibitem{latexcompanion}
Reig P., 2011, Astrophys. Space Sci., 332,1-29.
\bibitem{latexcompanion}
Reig P., Nespoli, E., 2013, A \& A, 551A, 1R
\bibitem{latexcompanion}
Sch{\"o}nherr G., Wilms J., Kretschmar P., Kreykenbohm I., Santangelo A., Rothschild R. E., Coburn W., Staubert R., 2007, A\& A,
472, 353
\bibitem{latexcompanion}
 Serim M. M. et al., 2022, MNRAS, 510, 1438–1449
\bibitem{latexcompanion}
Staubert R., 2014, Hercules X-1-another ’first’: long-term decay of the cyclotron line energy, preprint arXiv:1412.8067
\bibitem{latexcompanion}
Tsygankov, S. S., Lutovinov, A. A., Krivonos, R. R., Molkov, S. V., Jenke, P. J., Finger, M. H., Poutanen, J., 2016, MNRAS, 457,258
\bibitem{latexcompanion}
Vasco D., Staubert R,. Klochkov D., Santangelo A. et al. 2013, A\& A, 550, A111
\bibitem{latexcompanion}
Voges, W., Pietsch, W., Reppin, C., Trümper, J., Kendziorra, E., \& Staubert, R. 1982, ApJ, 263, 803 
\bibitem{latexcompanion}
Walter F., 1976, IAUC, 2959, 2
\bibitem{latexcompanion}
 Weng S. S. , Ge M. -Y. , Zhao H.-H., Wang W., Zhang S.-N., Bian W.-H. , Yuan Q.-R, 2017, ApJ, 843, 69
\bibitem{latexcompanion}
Wilson C. A., Finger M. H., Camero-Arranz A., 2008, ApJ, 678, 1263
\end{theunbibliography}
\end{document}